\documentclass[aps,amsmath,twocolumn,amssymb,floatfix,showpacs,superscriptaddress,nofootinbib,longbibliography]{revtex4-2}
\usepackage{braket}
\usepackage[dvipsnames]{xcolor}
\usepackage{float}
\usepackage{subfigure}
\usepackage{tikz}
\usepackage[colorlinks=true,linktoc=page,linkcolor=magenta,citecolor=Violet,urlcolor=purple]{hyperref}
\usepackage{amssymb}
\usepackage{amsfonts}

\mathchardef\mhyphen="2D 

\newcommand{\ie}{{ i.e.,\,\,}}
\newcommand{\eg}{{e.g.,~}}

\newcommand{\mc}{\mathcal}
\newcommand{\mbf}{\mathbf}

\newcommand\bea{\begin{eqnarray}}
\newcommand\eea{\end{eqnarray}}
\newcommand\beq{\begin{equation}}  
\newcommand\eeq{\end{equation}}

\usepackage{float}

\newcommand{\non}{\nonumber}  
\usepackage[normalem]{ulem}
\definecolor{lime}{HTML}{A6CE39}
\usepackage{sidecap,tikz}
\usepackage{amsmath} 
\DeclareRobustCommand{\orcidicon}{\hspace{-1.0mm}
	\begin{tikzpicture}
		\draw[lime, fill=lime] (0.0,0.0) 
		circle [radius=0.15] 
		node[white] {{\fontfamily{qag}\selectfont \tiny \,ID}};
		\draw[white, fill=white] (-0.0525,0.095) 
		circle [radius=0.007];
	\end{tikzpicture}
	\hspace{-3.0mm}
}
\foreach \x in {A, ..., Z}{\expandafter\xdef\csname orcid\x\endcsname{\noexpand\href{https://orcid.org/\csname orcidauthor\x\endcsname}{\noexpand\orcidicon}}
}

\AtBeginDocument{%
	\newwrite\bibnotes
	\def\bibnotesext{Notes.bib}
	\immediate\openout\bibnotes=\jobname\bibnotesext
	\immediate\write\bibnotes{@CONTROL{REVTEX41Control}}
	\immediate\write\bibnotes{@CONTROL{%
			apsrev41Control,author="08",editor="1",pages="1",title="1",year="1"}}
	\if@filesw
	\immediate\write\@auxout{\string\citation{apsrev41Control}}%
	\fi
}%

\begin{document}

\title{Current switching behavior mediated via hinge modes in higher-order topological phases using altermagnets}  

\author{Minakshi Subhadarshini\orcidA{}}\thanks{MS and AP contributed equally to this work.}
\affiliation{Institute of Physics, Sachivalaya Marg, Bhubaneswar-751005, India}
\affiliation{Homi Bhabha National Institute, Training School Complex, Anushakti Nagar, Mumbai 400094, India}
\author{Amartya Pal\orcidB{}}\thanks{MS and AP contributed equally to this work.}
\affiliation{Institute of Physics, Sachivalaya Marg, Bhubaneswar-751005, India}
\affiliation{Homi Bhabha National Institute, Training School Complex, Anushakti Nagar, Mumbai 400094, India}
\author{Arijit Saha\orcidD{}}
\email{arijit@iopb.res.in}
\affiliation{Institute of Physics, Sachivalaya Marg, Bhubaneswar-751005, India}
\affiliation{Homi Bhabha National Institute, Training School Complex, Anushakti Nagar, Mumbai 400094, India}


\begin{abstract}
We propose a theoretical framework to engineer hybrid-order and higher-order topological phases in three-dimensional topological insulators by coupling to $d$-wave altermagnets (AMs). Presence of only $d_{x^2-y^2}$-type AM drives the system into a hybrid-order topological phase where both first-order and second-order topological phases coexist. This phase is characterized by spectral analysis, low-energy surface theory, dipolar and quadrupolar winding numbers, and it's signature is further confirmed by two-terminal differential conductance calculations. Incorporation of the $d_{x^2-z^2}$-type AM drives the system into two second-order topological insulator phases hosting distinct type of hinge modes. These two variants of second-order topological phases are also topologically characterized by spectral analysis, topological invariants, low-energy surface thoery, and transport calculations. Importantly, the localization and direction of propagation of these one-dimensional hinge modes are controllable by tuning the relative strengths of the alermagnetic exchange orders. We utilize this feature to propose a tunable current-switching behaviour mediated via the hinge modes. Our results establish AMs based hybrid structure as a versatile platform for controllable higher-order topology and hinge-mediated device applications.
\end{abstract}

\maketitle

\section{Introduction}
 The advent of altermagnets (AMs)~\cite{Smejkal2022,Tomas2022,Mazin2023,Sayantika2024} has led to a surge of research activity in the condensed matter community. AMs represent a class of antiferromagnets with centro-symmetric collinear-compensated magnetic order. Unlike conventional ferromagnets, which exhibit uniform magnetization and spin-split energy bands, and unlike antiferromagnets with zero magnetization and no spin-splitting of energy bands, AMs feature both zero magnetization and momentum-dependent spin-split electronic bands.  Furthermore, AMs break the combined parity ($\mc{P}$) and time-reversal symmetry (TRS) ($\mc{T}$) giving rise to such momentum-dependent spin-splittng of energy bands \ie $E(\sigma,\mbf{k})\ne E(-\sigma,\mbf{k})$ ~\cite{Smejkal2022,Tomas2022,Mazin2023,Sayantika2024,Hayami_2019,Hayami_2020}. In $d$-wave AMs, opposite spin-sublattices are mapped onto each other by a $C_{4z}$ rotation rather than translation or reflection, which exhibits the net magnetization zero. Interestingly, $d$-wave AMs can be considered as a Zeeman field where the momentum-dependent magnetic order resembles the shape of the atomic planar $d$-orbitals instead of being uniform. This unique unconventional magnetic phase gives rise to a plethora of theoretical works \eg anomalous Hall effect~\cite{Sato2024_PRL,Zuzin2025,Sorn2025,Attias2024}, anisotropic spin-polarized conductivity~\cite{Dou_2025,Lin_2025}, anisotropic magnetoresistance~\cite{GonzalezBetancourt2024}, efficient spin–charge conversion ~\cite{Leivisk2024,Bai_2023}, topological superconductivity~\cite{Ghorashi2024_PRL,Li2024_HOTSC_AM,Zhu2023_TSC_AM,Maeda2025,Mondal_2025,Ohidul2025,pal2025AM_SDE}, Josephson effect~\cite{Ouassou2023,Lu2024_PRL,Sharma2025,Pal2025_PRBL,Sun2025,Fukaya_2025}, spintronics \cite{Chen_2025,Fu2025} etc. Various candidate materials with such anisotropic spin-polarized Fermi surfaces are proposed such as RuO$_2$~\cite{Bai_2023,Plouff2025}, MnF$_2$~\cite{Sayantika2024}, MnTe~\cite{Mazin2023} etc.

\begin{figure}[h]
	\centering
	\includegraphics[width=0.48\textwidth]{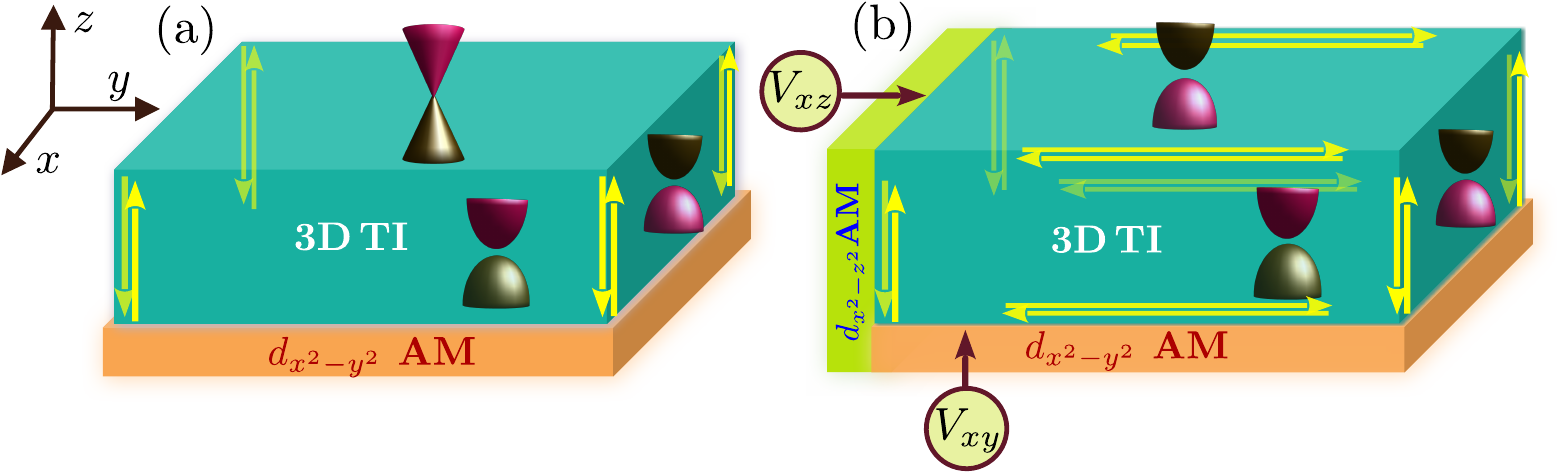}
	\caption{Schematic illustration of a 3D TI interfaced with AM layers. 
	(a) A 3D TI (cyan) is placed on top of a $d_{x^2-y^2}$ AM (orange) exhibiting gapless Dirac surface states on the top surface, while the surface states in the front and side surfaces are gapped.
	(b) Addition of a $d_{x^2-z^2}$ AM (green) at the $y=0$ plane of 3D TI gaps out all surface states and host only hinge modes. Two gate voltages $V_{xy}$ and $V_{xz}$ are assumed to be introduced at the interfaces located at the $z=0$ and $y=0$ planes, respectively, to control the proximity induced AM exchange strength. In both the panels, the propagating hinge modes are highlighted using the yellow arrows.}
	\label{Fig1}
\end{figure}
In recent times, the concept of higher-order topology generalizes the conventional bulk–boundary correspondence to boundaries of reduced dimensionality. While first-order topological insulators (TIs) host gapless states on $(d-1)$-dimensional surfaces~\cite{Fu_Kane_2007}, an $n^{\rm {th}}$-order topological insulator supports modes localized on $(d-n)$-dimensional boundaries, such as hinges or corners~\cite{Benalcazar2017,Benalcazar2017_1,Schindler2018,Ezawa2018,Franca2018,Vladimir2019,Trifunovic2019,Chatterjee2024}. These higher-order topological phases (HOTPs) ($n>1$) can arise from crystalline symmetries including reflection, rotation, and inversion~\cite{Felix2017,Song2017,Eslam2018,Jia_2024}, and have been realized in diverse material platforms such as metamaterials~\cite{Ni2019,Serra-Garcia2018}, bismuth-based compounds and transition-metal dichalcogenides~\cite{Schindler2018_HOTI,Zhang2020,Wang2019} etc. Also, HOTPs utilizing the $d$-wave AMs have been recently theoretically investigated in two dimensions (2D)~\cite{Ezawa2024_HOTI_AM,Cheng-Cheng_2024,Cheng_2023,Yang2025}.

Very recently, the idea of higher-order topology has been extended to hybrid-order topological phases (HyOTPs) that embody both first-order and second-order topological phases \cite{Ghorashi2020_HOWSM,Wang2020_HOWSM,Wang2019,Wang2020_HONLSM,Qiu2024,Pal_2025,Jia_2025,Yang_2025}. 
However, a systematic generation of HOTPs and HyOTPs based on altermagnetic platforms in 3D has not been investigated so far, to the best of our knowledge. Moreover, the exchange order in $d$-wave AMs closely mimics the Wilson-Dirac (WD) mass term, essential for realizing HOTPs~\cite{Schindler2018_HOTI,Arnob_2021}. However, WD mass preserves the combined $\mc{PT}$ symmetry, whereas AMs break $\mc{PT}$-symmetry. Thus, it would be really interesting to investigate the role of AMs in realizing HOTPs and HyOTPs in 3D. Given this background, in this article, we ask the following intriguing questions: (i) Can altermagnetic exchange generate both HOTPs and HyOTPs in 3D? (ii) If HyOTP and HOTP emerge, then how can they be distinctly identified in transport calculation? and (iii) Can higher-order boundary modes (1D hinge modes) be manipulated 
to yield an elegant, device-relevant transport functionality? 

To address these interesting questions, we propose a theoretical model consisting of $d$-wave AMs with $d_{x^2-y^2}$ and $d_{x^2-z^2}$ symmetry and a 3D TI as schematically shown in Fig.\,\ref{Fig1}. Our analysis identifies two distinct classes of topological phases: (i) a HyOTP with coexisting 2D surface and 1D hinge modes, (see Fig.\,\ref{Fig2}) and (ii) a purely second-order topological insulator (SOTI) phase hosting only 1D hinge-localized modes (see Fig.\,\ref{Fig4}). In addition to topological invariants such as dipolar winding number (DWN), quadrupolar winding number (QWN), and spectral analysis,  we also identify the HyOTP using transport calculation (see Fig.\,\ref{Fig3}). We also derive the effective low-energy surface theory to analytically understand the role of AMs order.  Importantly, we find the emergence of two types of SOTI phases where the localization and direction of propagation of hinge modes are controllable by the relative strengths of AMs. Utilizing this feature, we propose a current-switching like behaviour, where the conducting properties of the system can be controlled by tuning the strength of the AMs, which are consistent with both analytical predictions and numerical simulations. 



\section{Model} To begin with, we consider our system to be composed 
of a 3D first order TI placed on top of 2D $d$-wave AM with $d_{x^2-y^2}$ symmetry as schematically illustrated in  Fig.~\ref{Fig1}(a). The other layer with $d_{x^2-z^2}$ symmetry is attached at the $y=0$ surface of the 3D TI layer as shown in Fig.~\ref{Fig1}(b). This can be described by the following tight-binding Hamiltonian in the momentum space on a cubic lattice as~\cite{Zhang2009,Benalcazar2017,Tomas2022},
\begin{equation}
    \mc{H}(\mbf{k}) = \mc{H}^{\rm TI}(\mbf{k}) + \mc{H}^{\rm AM}_{xy}(\mbf{k}) + \mc{H}^{\rm AM}_{xz}(\mbf{k}) \label{eq:Hamiltonian}\ ,
\end{equation}
 with,
\begin{eqnarray}
		\mathcal{H}^{\rm TI}(\mathbf{k}) &=& 2\lambda (\sin k_x \Gamma_1 + \sin k_y \Gamma_2 + \sin k_z \Gamma_3)  \non \\
        		&& + [m_0 - 6t + 2t \sum_{i=x,y,z}\!\!\cos k_i\, ]\, \Gamma_4\ , \label{Eq:3D_TI_Ham}
\end{eqnarray}
\begin{eqnarray}
    	\mc{H}^{\rm AM}_{xy}(\mbf{k}) &=& J_{xy} (\cos k_x - \cos k_y)\, \Gamma_5\ , \\
	\mc{H}^{\rm AM}_{xz}(\mbf{k}) &=& J_{xz} (\cos k_z - \cos k_x)\, \Gamma_5\ .
\end{eqnarray}
where, $\mc{H}^{\rm TI}$, $\mc{H}^{\rm AM}_{xy}$, and $\mc{H}^{\rm AM}_{xz}$ represent the  3D TI, $d$-wave AMs having $d_{x^2-y^2}$, and $d_{x^2-z^2}$ symmetry, respectively. The  model parameters $\lambda$, $t$, and $m_0$ denote the strength of the intrinsic spin-orbit coupling (SOC), nearest neighbour hopping amplitude, and staggered mass term originating from crystal field splitting~\cite{Fu_Kane_2007}. The altermagnetic exchange order is assumed to be uniformly induced in the bulk of 3D TI via the proximity effect. 
In realistic systems, the proximity-induced altermagnetic exchange is confined near the interface and decays exponentially into the bulk (normal to the interface)~\cite{Wan_2025_1,Wan_2025_2,Chen2025_PRL}. However, our proposed results remain valid even under such a spatially varying exchange profile (see SM~\cite{supp} for further details).
Here, $J_{xy}$, $J_{xz}$ represent the strength of the altermagnetic exchange terms with $d_{x^2-y^2}$ and $d_{x^2-z^2}$ symmetry, respectively. The ($8\times 8$) $\Gamma$ matrices are chosen to form a set of anti-commutating matrices and defined as, $\Gamma_1=\mu_x\sigma_x s_y$, $\Gamma_2=\mu_z\sigma_0 s_0$, $\Gamma_3=\mu_x\sigma_z s_y$,
$\Gamma_4=\mu_x\sigma_y s_y$, $\Gamma_5=\mu_x\sigma_0 s_z$
where, $\mu_i$, $\sigma_i$, and $s_i$ ($i=0,x,y,z$) are Pauli matrices operating in the sublattice, orbital, and spin degrees of freedom. Note that, proximity induced altermagnetic exchange field in the TI need not to be diagonal in sublattice~\cite{Cheng_2023}. We assume an off-diagonal term, $\mu_x$~\cite{Cheng_2023}, as the minimal chiral-symmetry preserving coupling capturing inter-sublattice hybridization. Also, it satisfies the basic defintion (symmetry requirements) of $d$-wave AMs (see SM~\cite{supp} for details).

The Hamiltonian for the 3D TI, $\mc{H}^{\rm TI}(\mbf{k})$ (Eq.\,\eqref{Eq:3D_TI_Ham}) represents a first-order TI with gapless Dirac like surface states for $0<m_0<4t$~\cite{Fu_Kane_2007}. Also, $\mc{H}^{\rm TI}(\mbf{k})$ respects the TRS of the system with $\mc{T}=i\mu_x\sigma_0s_y\mc{K}$, $\mc{K}$ being the complex-conjugation operator: $\displaystyle{\mc{T}^{-1} \mc{H}^{\rm TI}(\mathbf{k}) \mc{T}=\mc{H}^{\rm TI}(-\mathbf{k})}$.  We refer the readers to supplementary material (SM)~\cite{supp} for further details on band structure and spatial distribution of the surface states. The altermagnetic Hamiltonians $\mc{H}^{\rm AM}_{xy}(\mbf{k})$ and $\mc{H}^{\rm AM}_{xz}(\mbf{k}$) individually break the TRS of the system and gaps out the surface states. 
In addition, $\mc{H}^{\rm AM}_{xy}(\mbf{k})$ breaks four fold rotation $(C_{4z})$ along $z$-axis but preserves the combined $C_{4z} \mathcal{T}$ symmetry. Similarly $\mc{H}^{\rm AM}_{xz}(\mbf{k})$ breaks four fold rotation $(C_{4y})$ along $y$-axis but preserves the combined $C_{4y} \mathcal{T}$ symmetry, thus satisfying the defining relations (symmetry requirements) of $d$-wave AMs. Specifically, the $d_{x^2 -y^2}$ type exchange gaps out the Dirac surface states in the $xz$ and $yz$ planes. While $d_{x^2 - z^2}$ symmetry is introduced to gap out the Dirac surface states in $xy$ plane to realise the SOTI phases, which we discuss in detail later. 
However, in principle, one can also consider AMs with $d_{xy}$ ($d_{xz}$) symmetry which can be mapped into the $d_{x^2-y^2}$ ($d_{x^2-z^2}$)-symmetry by rotating the coordinate axis via an angle of $\pi/4$ (see SM~\cite{supp} for details). Nevertheless, we consider AMs with only $d_{x^2-y^2}$ and $d_{x^2-z^2}$-symmetry for simplicity.



\section{Emergence of hybrid-order topology and its topological characterization} 
\begin{figure}
	\centering
	\includegraphics[width=0.5\textwidth]{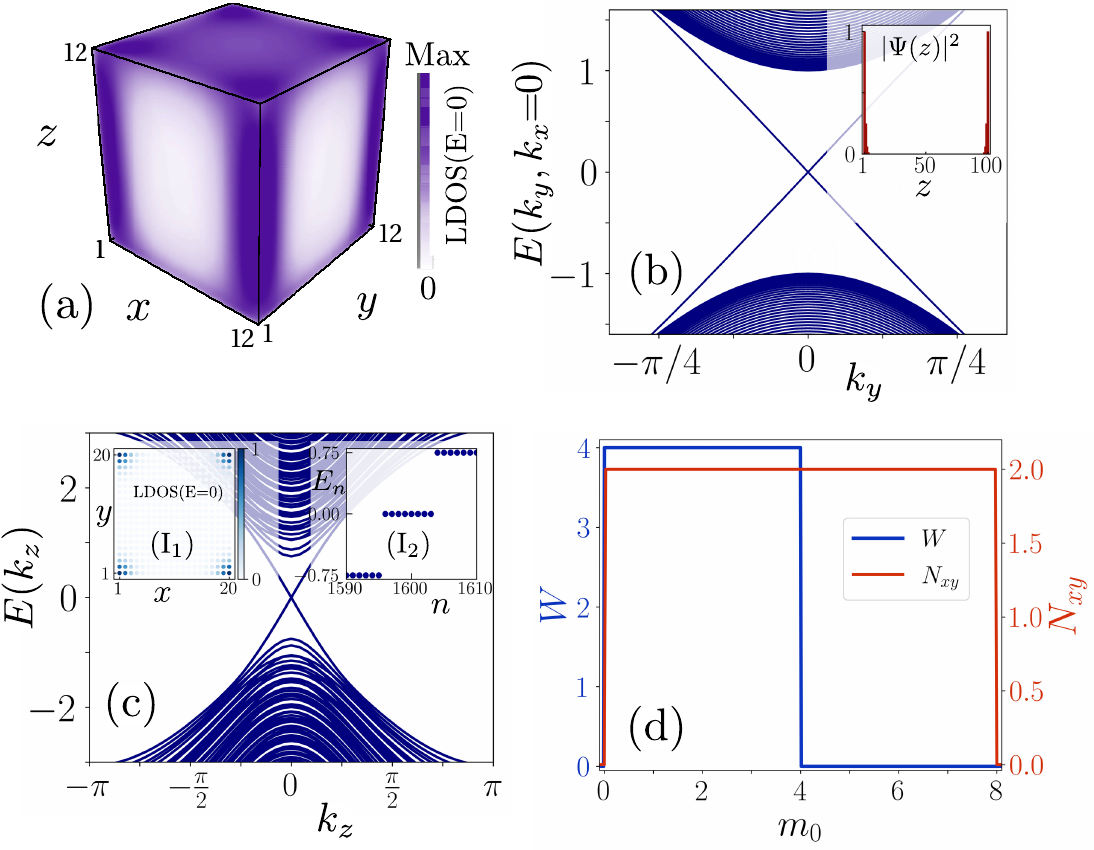}
	\caption{Spectral properties and topological invariants in the HyOTP. In panel (a), we show the LDOS ($E=0$) in the HyOTP considering a finite geometry along the $x,y,$ and $z$ directions with $L_x=L_y=L_z=12$ lattice sites. Panel (b) displays the energy spectrum, corresponding to finite system size along $z$ directions with $L_{z}=100$, as a function of $k_y$ with $k_x=0$, while the inset exhibits the spatial profile $|\psi(z)|^2$ at $k_x=k_y=0$. Panel (c) shows the hinge spectrum as a function of $k_z$, with insets ($I_1$, $I_2$) illustrating the LDOS at $E=0$ and the eigenvalue spectrum with respect to eigenstate index $n$ with $L_x=L_y=20$. In panel (d), we present the topological invariants, DWN $W$ (left axis) and QWN $N_{xy}$ (right axis) as a function of $m_0$. Other model parameters ae chosen as, $m_0 = t$, $t=1$, $\lambda=t$, $J_{xy}=2t$ and $J_{xz}=0$.}
	\label{Fig2}
\end{figure}

We first discuss the effect of only $\mc{H}^{\rm AM}_{xy}(\mbf{k})$ on the surface states of 3D TI. We systematically study the spectral features like band structure, local density of states (LDOS) to probe the bulk-boundary correspondence, and topological invariants to characterize the HyOTP. First, we compute the site-resolved LDOS of the zero-energy eigenstates of $\mc{H}(\mbf{k})$ by diagonalizing the Hamiltonian in real space with a finite system size $L_x,L_y,$ and $L_z$ along the $x,y$, and $z$ directions employing open boundary conditions (OBC). The corresponding LDOS ($E=0$) is depicted in the Fig.\,\ref{Fig2}(a) which indicates the presence of both surface- and hinge-localized boundary states. The surface states appear in the $x-y$ plane ($z=0,L_z$) while the hinge modes appear perpendicular to the $x-y$ plane \ie along $z$-direction. To identify these surface states, we diagonalize $\mc{H}(\mbf{k})$ in slab geometry (see SM~\cite{supp} for details) and display the band structure in Fig.\,\ref{Fig2}(b) as a function of $k_y$ with $k_x=0$. This exhibits the presence of gapless states. From the LDOS ($E=0$) behavior (inset of Fig.\,\ref{Fig2}(b)) we find the gapless states to be localized around $z=0$ and $z=L_z$, confirming them to be surface states. 

Then, in order to identify the hinge localized modes, we compute the energy-eigenvalues of the Hamiltonian $\mc{H}(\mbf{k})$ considering rod geometry (see SM~\cite{supp} for details) and depict as a function of $k_z$ in the Fig.\,\ref{Fig2}(c). We observe the emergence of gapless dispersive states, crossing each other at $k_z=0$. Furthermore, we calculate the energy eigenvalues, $E_n$, and LDOS ($E=0$) by setting $k_z=0$. We depict the variation of $E_n$ as a function 
of state index, $n$, in the inset (${\rm I_1}$) 
and the LDOS ($E=0$) in the inset $(\rm I_2)$ of Fig.\,\ref{Fig2}(c) respectively. From these plots, we infer the appearence of a total of eight gapless hinge-localized states establishing the emergence of SOTI in 3D.  

To this end, we topologically characterize the HyOTP by computing the DWN $W$, and QWN $N_{xy}$ to identify the first and second order topological phases, respectively.
The DWN, used to characterize a gapped 1D topological phase is defined as~\cite{Chiu_2016,Mondal_2025,Pal_2025},
\begin{equation}
	W = \frac{i}{2\pi} \int_{\text{BZ}} dk_z\, \text{Tr}\!\left[h^{-1}(k_z)\,\partial_{k_z} h(k_z)\right]\ ,
\end{equation}
where, \( h(k) \) can be obtained by antidiagonalizing \( \mathcal{H}({\mbf{k}} )\) of Eq.~(\ref{eq:Hamiltonian}) in the chiral basis with chiral operator $ \mc{S} = \mu_0\sigma_x s_0 $.  We obtain an effective 1D system by setting $k_x=k_y=0$ in $\mc{H} ({\mbf{k}})$ since the surface states intersect with each other at $k_x=k_y=0$ (see Fig.\,\ref{Fig2}(b)). We refer the readers to SM~\cite{supp} for further details.   





On the other hand, QWN is utilized to characterize the SOTIs in 2D, which we obtain by setting $k_z=0$ in $\mc{H}({\mbf{k}})$ since the hinges modes intersect at $k_z=0$. Exploiting the chiral symmetry ($\mc{S}=\mu_y\sigma_zs_z$) in this geometry, the QWN  can be defined as as~\cite{Benalcazar_2022,Pal_2025,Subhadarshini_2025},
\begin{align}
    N_{xy} = \frac{1}{2\pi i}\mathrm{Tr} \ln \left(\bar{Q}_A \bar{Q}_{B}^\dagger\right)  \ ,
    \label{eq:Quad}
\end{align}
where, $\bar{Q}_{A,B}$ stands for sublattice quadrupole operator, defined as $\bar{Q}_{A,B}=U_{A,B}^\dagger U^S_{A,B} Q U^S_{A,B} U_{A,B}^\dagger$ with the quadrupole operator given by, $Q=\exp \left(-2\pi i xy/L_xL_y\right)$. The $U^S_{A(B)}$ and $U_{A(B)}$ are unitary matrices obtained from the chiral symmetry operator and anti-diagonalizing the $\mc{H}({\mbf{k}})$ (see SM~\cite{supp} for further details).

We discuss the variation of $W$ (left axis) and $N_{xy}$ (right axis) as a function of the staggered mass term $m_0$ in Fig.\,\ref{Fig2}(d). For $0\le m_0 \le 4t$, $W$ is nonzero while zero outside this range, which still corresponds to the topological regime of the parent 3D TI~\cite{Fu_Kane_2007}. In contrast, $N_{xy}=2$ for $0\le m_0 \le 8t$, which indicates the presence of SOTI phase with two modes per hinge and validates our results as shown in Fig.\,\ref{Fig2}(c). Thus, for $0<m_0<4t$, both $W$ and $N_{xy}$ are nonzero which confirm the presence of HyOTP comprised of both first and second order topological phases. Importantly, outside the topological regimes of the 3D TI \ie $4t\le m_0\le 8t$, $N_{xy}=2$ and $W=0$, leading to the emergence of a pure SOTI phase. Therefore, the presence of AM also extends the topological phase boundary beyond that of the first-order 3D TI. 

For a more fundamental understanding of the role of AM in realizing the HyOTP, we analytically derive the low-energy effective Hamiltonians for $xy$, $yz$, and $xz$ surfaces utilizing the low-energy surface theory as,~\cite{Arnob_2021,Subhadarshini_2025} (for details see SM~\cite{supp}):
\begin{equation}
	H_{xy}^S = -2\lambda k_x \sigma_x s_y - 2\lambda k_y \sigma_z s_0\ , \label{Eq.H_xy_S}
\end{equation}
\vskip -0.7cm
\begin{equation}
		H_{yz}^S = -2\lambda k_y \sigma_z s_0 + 2\lambda k_z \sigma_x s_y
		+ \frac{J_{xy} m_0}{2t}\sigma_x s_z \ , 
\end{equation}
\vskip -0.7cm
\begin{equation}
		H_{xz}^S = -2\lambda k_x \sigma_z s_0 + 2\lambda k_z \sigma_x s_0 
		+ \frac{J_{xy} m_0}{2t}\sigma_y s_x \ .
\end{equation}
From these effective Hamiltonians, it is evident that $xy$-surface states do not acquire any mass term, reflecting the gapless surface state as the signature of first-order topology, while the $yz$ and $xz$ surfaces become gapped due to the acquired mass terms ($\propto J_{xy}m_{0}$). To compare the induced mass terms in the $yz$ and $xz$ surface, we perform a unitary transformation in $H_{xz}$ and show that $xz$- and $yz$-surfaces acquire equal and opposite masses, $\pm (J_{xy} m_0 / 2t)$ (see SM~\cite{supp} for details). This creates a mass-domain wall configuration. Such a domain wall configuration supports zero-energy Jackiw-Rebby modes which reflect as hinge modes in the SOTI phase in 3D system~\cite{Jackiw_1976}.

\section{Transport characteristics of the H{\lowercase{y}}OTP} 
\begin{figure}
	\centering
	\includegraphics[width=0.49\textwidth]{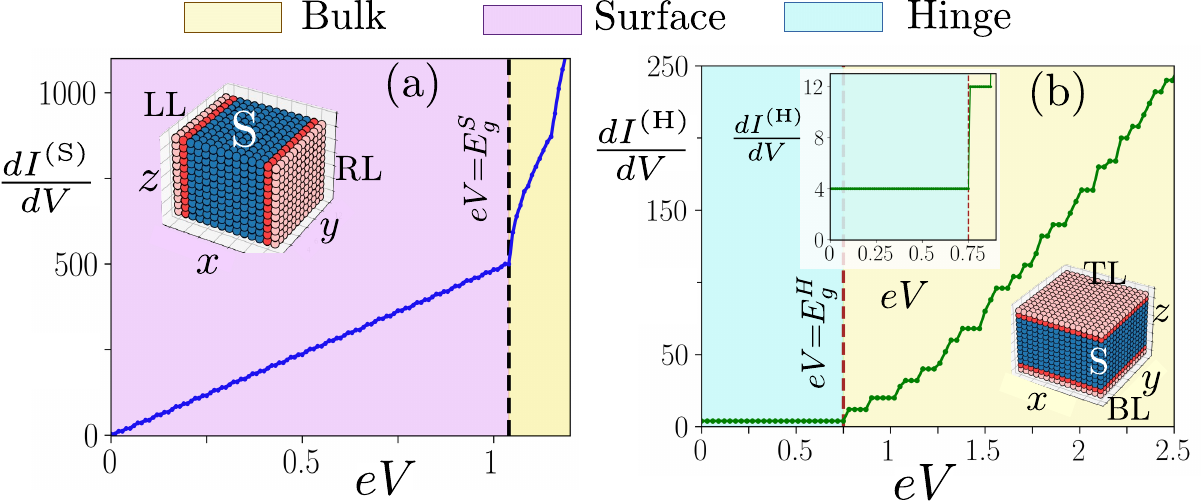}
	\caption{Transport characteristics in the HyOTP. In panel (a), we depict the differential conductance $\frac{dI^{\rm (S)}}{dV}$ as a function of $(eV)$, due to the surface states in HyOTP. The inset illustrates the transport setup with LL (left lead), RL (right lead), and S (system) through which the current flows. Panel (b) presents the differential conductance $\frac{dI^{\rm (H)}}{dV}$ with respect to $(eV)$ for the hinge states. The inset (bottom) highlights the transport setup with top and bottom leads (TL and BL). On the other hand, the inset (top) emphasizes the $4e^{2}/h$ contribution arising due to hinge modes of the main plot. For panel (a) $L_{z}=100$ and panel (b) $L_{x}=L_{y}=20$. Other model parameters remain same as mentioned in Fig.~\ref{Fig2}. 
	}
	\label{Fig3}
\end{figure}

After discussing the spectral features in details, we now focus on the transport properties of the HyOTP. In particular, we compute the two-terminal differential conductance ($dI/dV$) within the  Landauer--B\"uttiker
 formalism~\cite{datta_1995} utilizing the Python package Kwant~\cite{Groth_2014}.  The gapless Dirac surface states, located at $z=0$ and $z=L_z$, disperse linearly with momentum along $x$- and $y$-directions (see Eq.\,\eqref{Eq.H_xy_S}). Thus, without loss of generality, we attach two semi-infinite leads 
at $x=0$ (left lead (LL)) and $x=L_x$ (right lead (RL)) as shown in the inset of Fig.\,\ref{Fig3}(a). We compute the differential conductance $dI/dV$ in this geometry in the presence of a voltage bias, $eV$, at zero temperature as, ~\cite{datta_1995,Groth_2014,Pal2024_WSM,Pal_2025}, 
\begin{equation}
    \frac{dI^{\rm (S)}}{dV}(eV) = \frac{e^2}{h} \sum_{k_y} {\rm Tr} \left[ t_S^\dagger(E) t_S(E)\right]_{E=eV} \ ,  \label{Eq.dIdV_S}
\end{equation}
Here, $t_S(E)$ denotes the transmission amplitude matrix for the geometry under consideration. The summation $\sum_{k_y}$ takes into account the propagating modes along the $y$-direction (see SM~\cite{supp} for the details). 

To identify the dispersive hinge modes propagating along the $\pm z$ direction (as illustrated in Fig.\,\ref{Fig2}(c)), we compute the differential conductance by attaching two leads at $z=0$ (bottom lead (BL)) and $z=L_z$ (top lead, (TL)) with a finite size system $L_x$ and $L_y$ along $x$ and $y$ directions as shown in the inset of Fig.\,\ref{Fig3}(b). The differential conductance in this geometry can be obtained as~\cite{datta_1995,Groth_2014,Pal2024_WSM,Pal_2025},
\begin{equation}
    \frac{dI^{\rm (H)}}{dV}(eV) = \frac{e^2}{h}  {\rm Tr} \left[ t_H^\dagger(E) t_H(E)\right]_{E=eV}\ , \label{Eq.dIdV_H}
\end{equation}
where, the transmission matrix `$t_H(E)$' incorporates all the information about the transverse modes (see SM~\cite{supp} for details). Note that, the leads are considered to
be of the same material as the system for simplicity.

We showcase the behavior of $\frac{dI}{dV}^{\rm (S)}(eV)$ and $\frac{dI}{dV}^{\rm (H)}(eV)$ (in units of $e^2/h$) as a function of $eV$ in Fig.\,\ref{Fig3}(a) and Fig.\,\ref{Fig3}(b), respectively. In Fig.\,\ref{Fig3}(a), $\frac{dI}{dV}^{\rm (S)}$ initially grows linearly with $eV$ within the bulk gap $E=E_g^S$. The reason can be attributed to the fact tha the gapless Dirac surface states exist only within the bulk gap and disperse linearly with momentum $k_x,k_y$ around the Dirac point. This leads to linearly varying density of states in 2D \ie ${\rm DOS}(E)\sim E$ for $E\le E_g^S$, which is being reflected in the transport calculation as a linear dependency with $eV$. For $E>E_g^S$, $\frac{dI}{dV}^{\rm (S)}$ deviates from the linear variation as it encounters the density of states due to the bands of the Dirac cone, which is not linear in $E$.

In case of $\frac{dI}{dV}^{\rm (H)}(eV)$, we obtain a quantized conductance of $4e^2/h$ 
for $eV\le E_g^H$ as depicted in Fig.\,\ref{Fig3}(b) and it's inset for better clarity. This quantization arises because the system hosts eight hinge modes in total, four propagating along $+z$ and four along $-z$ direction. Since the voltage bias $eV$ is applied across the top and bottom lead (TL and BL), $\frac{dI}{dV}^{\rm (H)}$ captures the signature of hinge modes propagating only along the $-z$ direction, leading to quantized conductance of $4e^2/h$. While such quantization is absent for the bulk states, $E>E_g^H$ and $\frac{dI}{dV}^{\rm (H)}$ becomes continuous. These results confirm that the HyOTP contains both surface and hinge-localized modes, as clearly demonstrated in the transport signatures.


\section{Emergence of SOTI and its implications in current-switching behavior via hinge modes} 
\begin{figure}
	\includegraphics[width=0.5\textwidth]{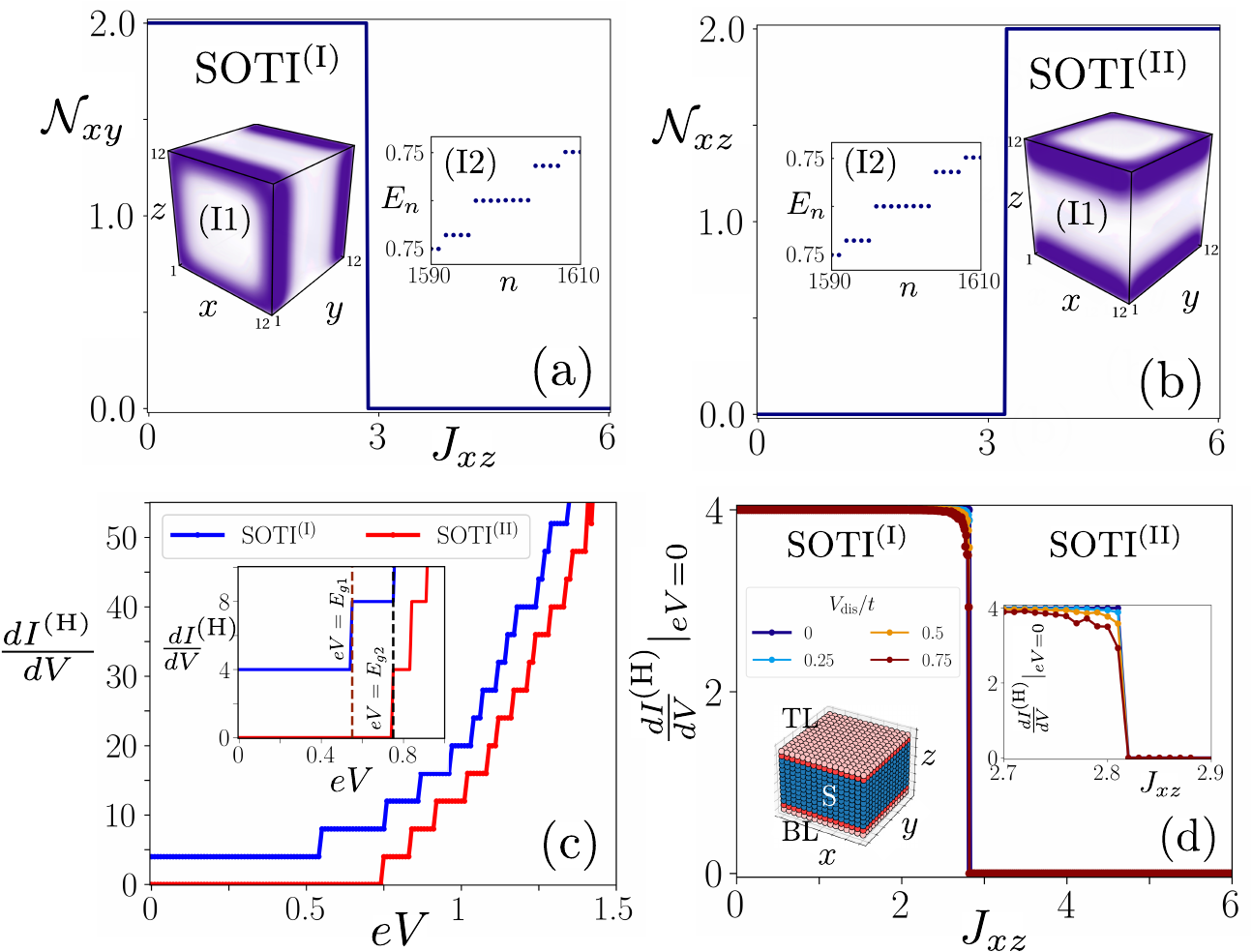}
	\caption{Topological characterization and current switching behavior in the SOTI phase. We depict $N_{xy}$ and $N_{xz}$ for the $xy$- and $xz$-planes in panels (a) 
		and (b), corresponding to SOTI$^{\text{(I)}}$ and SOTI$^{\text{(II)}}$ phases respectively. Insets (I1) and (I2) illustrate the eigenvalue spectrum $E_{n}$ with respect to $n$. The  LDOS distribution at $E=0$ for both the cases are also shown in panels (a) and (b). Panel (c):  $dI^{\rm (H)}/dV$ is displayed as a function of bias $eV$ for the SOTI$^{\rm (I)}$ ($J_{xz}=2t$) and SOTI$^{\rm (II)}$ phase ($J_{xz}=4t$); the inset exhibits an expanded view of the same conductance plot to highlight the contribution arising from the hinge modes. (d) Variation of $\frac{dI^{\rm (H)}}{dV}|_{eV=0}$ is displayed as a function of $J_{xz}$ for various disorder strengths $V_{\rm dis}$ after averaging over 30 independent disorder configurations. In the inset, $\frac{dI^{\rm (H)}}{dV}|_{eV=0}$ close to the phase transition point is highlighted for better clarity. Other model parameters are chosen as: $(m_0,\lambda,J_{xy})=(t,t,3t)$ and $(L_x,L_y,L_z)=(20, 20,10)$.}
	\label{Fig4}
\end{figure}

We here incorporate the AM with $d_{x^2-z^2}$-symmetry and systematically analyse the spectral and transport properties. We establish the emergence of two types of SOTI phase with 1D hinge modes. Interestingly, the localization and direction of propagation of these hinge modes can be controlled by tuning the strength of $J_{xz}$ relative to $J_{xy}$. We first derive the low energy-effective Hamiltonian in case of  both $J_{xy}, J_{xz}\ne0$, and obtain the induced mass terms on each surface with the following dependency on $J_{xy}$ and $J_{xz}$ as, $xy$-surface: $\displaystyle{\frac{J_{xz}m_0}{2t}}$, $yz$-surface: $\displaystyle{\frac{(J_{xy}-J_{xz})m_0}{2t}}$, and $xz$-surface: $-\displaystyle{\frac{J_{xy}m_0}{2t}}$ (see SM~\cite{supp} for details). Interestingly, the $xy$ surface now acquires a finite mass gap due to finite $J_{xz}$. However, the sign of mass term on the $yz$ surface depends on the relative strength of $J_{xy}$ and $J_{xz}$, which leads to the emergence of two types of SOTI phases, (i) SOTI$^{\rm (I)}$ for $J_{xz}<J_{xy}$ (see LDOS (E=0) plot in the inset of Fig.\,\ref{Fig4}(a)), and (ii) SOTI$^{\rm (II)}$ for $J_{xz}>J_{xy}$ (see LDOS (E=0) behavior in the inset of Fig.\,\ref{Fig4}(b)). Note that, SOTI$^{\rm (I)}$ phase consists of eight hinge modes propagating along the $x$ and $z$ directions, while SOTI$^{\rm (II)}$ phase anchors eight hinge modes propagating along the $x$ and $y$ directions. We topologically characterize the SOTI$^{\rm (I)}$ and SOTI$^{\rm (II)}$ phases by computing QWNs $N_{xy}$ and $N_{xz}$, respectively. The QWN $N_{xz}$ is computed similarly as $N_{xy}$ (as outlined in Eq.\,\eqref{eq:Quad}) with chiral symmetry $\mc{S}=\mu_0\sigma_0 s_x$ (see SM~\cite{supp} for further details). In particular, $N_{xy}$ ($N_{xz}$) characterize the hinge modes which are localized in the $xy$ ($xz$) plane and propagate along the $z$ ($y$) direction. We depict the variation of $N_{xy}$ and $N_{xz}$ by varying the strength of $J_{xz}$ with fixed $J_{xy}=3t$ in  Fig.\,\ref{Fig4}(a) and Fig.\,\ref{Fig4}(b), respectively. We find that $(N_{xy},N_{xz})=(2,0)$ for $J_{xz} <J_{xy}$ which corresponds to the SOTI$^{\rm (I)}$ phase. On the other hand, $(N_{xy},N_{xz})=(0,2)$ for $J_{xz}>J_{xy}$ corresponds to the SOTI$^{\rm (II)}$ phase. 

To identify the two types of SOTI phases in transport, we consider a setup where two leads are attached at $z=0$ (BL) and at $z=L_z$ (TL) (see the inset of Fig.\,\ref{Fig4}(d) for the setup), with a finite size system along the $x$ and $y$ directions. A voltage bias, $eV$, is applied across the top and bottom leads to compute the differential conductance resulting from the hinge modes propagating along the $z$-direction. We compute and depict the corresponding $\frac{dI}{dV}^{\rm (H)}(eV)$ (Eq.\,\eqref{Eq.dIdV_H}) in Fig.\,\ref{Fig4}(c) 
in case of both the SOTI$^{\rm (I)}$ and SOTI$^{\rm (II)}$ phases. 
Moreover, in the inset of Fig.\,\ref{Fig4}(c), we mark the bulk gaps, $E_{g1}$ (in the SOTI$^{\rm (I)}$ phase) and $E_{g2}$ (in case of SOTI$^{\rm (II)}$ phase) to explicitly highlight the contribution to the conductance arising from the hinge states ($eV<E_{g1},E_{g2}$) and bulk states ($eV>E_{g1}, E_{g2}$). Interestingly, we observe that  $\frac{dI}{dV}^{\rm (H)}$ attains a quantized value of $4e^2/h$ for $E\le E_{g1}$ only in the SOTI$^{\rm (I)}$ phase, while $\frac{dI}{dV}^{\rm (H)}$ remains zero for $eV\le E_{g2}$ in case of SOTI$^{\rm (II)}$ phase. The reason can be attributed to the fact that in the SOTI$^{\rm (I)}$ phase, four gapless dispersive hinge modes are propagating along the $z$-direction, which results in a total conductance of $4e^2/h$. However, such hinge modes are absent in the SOTI$^{\rm (II)}$ phase. We extend our analysis by incorporating onsite random static disorder of the form, $\mc{H}_{\rm dis} = \sum_{\mbf{r}} V_{\mbf{r}} \,\mu_0\sigma_0s_0$, where $V_{\mbf{r}}$ is a random variable drawn from a box distribution $[ -V_{\rm dis}, V_{\rm dis} ]$ and corresponds to the onsite potential at $\mbf{r}=(x,y,z)$. We then compute the zero-bias conductance, $\frac{dI}{dV}^{\rm (H)}|_{eV=0}$, by varying the strength of $J_{xy}$ and present the results in Fig.\,\ref{Fig4}(d) choosing different disorder strengths. Notably, in the clean limit ($V_{\rm dis}=0$), we observe that $\frac{dI}{dV}^{\rm (H)}|_{eV=0}$ becomes non-zero and quantized when $J_{xz}<J_{xy}$ while it is zero if $J_{xz}>J_{xy}$. This is due to the fact that the gapless hinge modes propagating along the $z$-direction exist only in the SOTI$^{\rm (I)}$ phase \ie for $J_{xz}<J_{xy}$ (see Fig.\,\ref{Fig4}(a)), which leads to the conductance quantization. Whereas, for $J_{xz}>J_{xy}$, the system enters into the SOTI$^{\rm (II)}$ phase where hinge modes propagate only along the $x$ and $y$ direction, not along the $z$-direction (see Fig.\,\ref{Fig4}(b)). This special feature can be utilized as a current-switching mechanism mediated via the 1D hinge modes where, tuning the strength of $J_{xz}$ relative to $J_{xy}$, one can control the direction of current flow along a particular direction. Importantly, we also observe that the current-switcing behavior survives even for finite disorder strengths as shown in the inset of Fig.\,\ref{Fig4}(d) (see SM~\cite{supp} for more details). We also examine the current-switching behavior with an exponentially decaying profile of altermagnetic exchange inside TI. In this case, while exact conductance quantization is lifted, the switching effect still remains clearly visible (see SM~\cite{supp} for further details). However, by choosing appropriate lead configuration, one can also make  SOTI$^{\rm (II)}$ phase conducting in this switching behavior. 
This feature might carry high relevance and utility in modern electronic devices. 

\section{Summary and Conclusion} To summarize, in this article, we put forward a theoretical framework for realizing both HOTPs and HyOTPs in a 3D TI proximitized with $d$-wave AMs with $d_{x^{2}-y^{2}}$ and $d_{x^{2}-z^{2}}$ symmetries. Coupling only to the $d_{x^{2}-y^{2}}$ AM drives the system into a HyOTP with gapless 2D Dirac surface states (first order) coexisting with the 1D hinge-localized modes (second order). Through a systematic analysis of spectral properties together with dipolar and quadrupolar winding numbers, we establish the existence of this HyOTP. Employing a low-energy surface theory, we analytically derive the effective Hamiltonians for each surface, clarifying how the AM exchange fields gap out selected surface states resulting in appearance of hinge modes. Finally, computing differential conductance within the Landauer--B\"uttiker formalism~\cite{datta_1995}, we identify both surface and hinge transport channels, providing clear signatures of the hybrid-order topology.

In presence of both $d_{x^{2}-y^{2}}$ and $d_{x^{2}-z^{2}}$ AMs the system hosts two distinct SOTI phases,  SOTI$^{\rm (I)}$ and SOTI$^{\rm (II)}$ where the localization and direction of propagation of the hinge modes are controlled solely by the relative strengths of the two AM components. Utilizing this feature, 
we demonstrate a hinge-mediated current-switching mechanism. For a given transport setup, SOTI$^{\rm (I)}$ phase exhibits nonzero quantized $dI^{\rm (H)}/dV(eV=0)$, whereas SOTI$^{\rm (II)}$ phase do not conduct. Thus, by simply tuning the AM exchange strengths, one can control the hinge-mode transport, effectively realizing a topological current switch with possible potential device relevance. Note that, an optimal thickness of the 3D TI (a few quintuple layers) is required to prevent hybridization of the top and bottom Dirac surface states while ensuring that the proximity-induced altermagnetic exchange does not decay too rapidly into the bulk. As possible materials, one may use the heterostructure of  $\rm{Bi_{2}Se_{3}}$-$\rm{MnTe}$, exhibiting room-temperature altermagnetism with an energy scale upto 370 meV~\cite{Zhang2009,Lee2024_MnTe}. Importantly, in proximity-coupled heterostructures, the strength of magnetic exchange field can be tuned via electrostatic gating,~\cite{Zhang2012,Lazi2016,Hellman2017,Katmis2016,Jiang2018,Bobkov2024}. As shown in Fig.~\ref{Fig1}(b), we consider two such independent gate voltages $V_{xy}$ and $V_{xz}$ to tune the relative strengths of the $d_{x^2-y^2}$ and $d_{x^2-z^2}$ AM layers proximitized to the 3D TI. By keeping one gate voltage fixed and varying the other, the relative strength of the AM exchange fields can be tuned. This allows possible experimental control over the relative strengths of $J_{xy}$ and $J_{xz}$ and thereby enabling the proposed switching effect.

So far, our discussion has centered around how the interplay between a first order 3D TI and 2D $d$-wave AMs gives rise to second-order topological phase in 3D. We now propose an additional route to realize a third-order TI (TOTI) phase in 3D by introducing an unconventional magnetic order that mimics the spatial profile of a $d_{z^{2}}$ orbital, $ \mathcal{H}_{d_{z^{2}}}(\mathbf{k}) \sim 2\cos k_z - \cos k_x - \cos k_y$. In the SM~\cite{supp}, we present a detailed analysis of this TOTI phase, including band structure, LDOS, topological invariant, and an analytic derivation of the effective hinge Hamiltonians supporting the numerical findings. We emphasize that such a $d_{z^{2}}$-type magnetic order is purely theoretical and currently lacks any material realization reported in literature. 
\section*{Acknowledgments} We acknowledge the Department of Atomic Energy (DAE), Govt. of India for providing the financial support. We also acknowledge SAMKHYA: High-Performance Computing Facility provided by Institute of Physics, Bhubaneswar and the two workstations provided by the Institute of Physics, Bhubaneswar from the DAE APEX project for numerical computations.

\section*{Data Availibility Statement} The data that support the findings of this article are not
publicly available. The data are available from the authors upon reasonable request.

\bibliographystyle{apsrev4-2mod}
\bibliography{bibfile}


\normalsize\clearpage
\begin{onecolumngrid}
	\begin{center}
		{\fontsize{12}{12}\selectfont
			\textbf{Supplementary Material for ``Current switching behaviour mediated via hinge modes in higher order topological phase using altermagnets''\\[5mm]}}
		{\normalsize  Minakshi Subhadarshini\orcidA{},$^{1,2,*}$Amartya Pal\orcidB{},$^{1,2,*}$  and  Arijit Saha\orcidD{},$^{1,2}$ \\[1mm]}
		{\small $^1$\textit{Institute of Physics, Sachivalaya Marg, Bhubaneswar-751005, India}\\[0.5mm]}
		{\small $^2$\textit{Homi Bhabha National Institute, Training School Complex, Anushakti Nagar, Mumbai 400094, India}\\[0.5mm]}
		{}
	\end{center}
	
	\newcounter{defcounter}
	\setcounter{defcounter}{0}
	\setcounter{equation}{0}
	\renewcommand{\theequation}{S\arabic{equation}}
	\setcounter{figure}{0}
	\renewcommand{\thefigure}{S\arabic{figure}}
	\setcounter{page}{1}
	\pagenumbering{roman}
	
	\renewcommand{\thesection}{S\arabic{section}}

%
%

%
%

\newcounter{specialsection}
\newcounter{specialsubsection}[specialsection]
\renewcommand{\thespecialsection}{S\arabic{specialsection}}
\renewcommand{\thespecialsubsection}
{\thespecialsection.\arabic{specialsubsection}}

\newcommand{\specialsection}[1]{
	\refstepcounter{specialsection}
	
	\addcontentsline{stoc}{section}
	{\protect\numberline{\thespecialsection}#1}
	
	\bigskip
	\noindent
	{\normalsize\bfseries
		\thespecialsection\quad #1}
	\par\medskip
}

\newcommand{\specialsubsection}[1]{
	\refstepcounter{specialsubsection}
	
	\addcontentsline{stoc}{subsection}
	{\protect\numberline{\thespecialsubsection}#1}
	
	\medskip
	\noindent
	{\bfseries
		\thespecialsubsection #1 \quad}
	\par\smallskip
}

\section*{Contents}

\makeatletter
\@starttoc{stoc}
\makeatother
\vskip 9.5cm
	{*MS and AP contributed equally to this work.}

	\newpage
	
	\begin{center}
		\specialsection{\hskip 0.4cm Relation between altermagnets (AMs) with $d_{x^2-y^2}$ and $d_{xy}$ symmetry} \label{Sec:I}
	\end{center}		

	In the main text, we have considered altermagnetic exchange order with $d_{x^2-y^2}$ symmetry and mentioned that $d_{x^2-y^2}$ AM is related to $d_{xy}$ AM with a coordinate transformation. Here, we show that explicitly. In two dimensions (2D), the $d_{xy}$ altermagnetic term can be written as~\cite{Tomas2022}
	\[
	H_{2D}(k_x,k_y) = J_{xy} k_x k_y\ .
	\]
	By parameterizing the momentum as $k_x = k_F \cos\theta$ and $k_y = k_F\sin\theta$ ($k_F$ denotes the Fermi momentum), we obtain
	\begin{equation}
		\begin{aligned}
			J_{xy} k_x k_y &= J_{xy}k_F^2 \sin\theta \cos\theta = \frac{J_{xy}k_F^2}{2}\sin(2\theta)\ .
		\end{aligned}
	\end{equation}
	
	Then, we perform a rotation of the coordinate system by $\pi/4$, \ie $\theta \rightarrow \theta' - \pi/4$. Hence,
	\begin{equation}
		\begin{aligned}
			\frac{J_{xy}k_F^2}{2}\sin(2\theta)
			&= \frac{J_{xy}k_F^2}{2}\sin(2\theta' -\pi/2) \\
			&= \frac{J_{xy}k_F^2}{2}\cos(2\theta') \\
			&= \frac{J_{xy}k_F^2}{2}(\cos^2\theta' - \sin^2\theta') \\
			&= \frac{J_{xy}}{2}(k_y'^2 - k_x'^2).
		\end{aligned}
		\label{Altermag_trans}
	\end{equation}
	Thus, rotating the frame of reference by an angle $\pi/4$, the $J_{xy}k_xk_y$ term transforms to a form proportional to $J_{xy}(k_y^2 - k_x^2)$.
	
	Similarly, the other altermagnetic term transforms in the similar manner as:
	\[
	J_{xz} k_x k_z \ \longrightarrow \ \frac{J_{xz}}{2}(k_x^2 - k_z^2).
	\]  
	
	Thus, for simplicity we have only considered the AMs with $d_{x^2-y^2}$ type exchange order parameter in the main text.

		\begin{center}
		\specialsection{\hskip 0.4cm Hamiltonian and band structure in various topological phases}\label{Sec:II}
	\end{center}
	
	In the main text, we have emphasized the topological invariants, spectral properties, and transport signatures in hybrid-order and second-order topological phases considering various finite size geometries. In this section, we explicitly derive the Hamiltonians in these finite geometries namely slab, rod, and cubic geometries and also present the corresponding band structures. The latter is not shown in the main text. We begin by writing the momentum space tight-binding Hamiltonian (as mentioned in Eq.\,(1)-Eq.\,(4) of the main text) as~\cite{Zhang2009,Benalcazar2017,Tomas2022}, 
	%
	%
	\begin{equation}
		\begin{aligned}
			\mathcal{H} = &2\lambda (\sin k_x \Gamma_1 + \sin k_y \Gamma_2 + \sin k_z \Gamma_3)  + [m_0 - 6t + 2t (\cos k_x + \cos k_y + \cos k_z)] \Gamma_4\\ 
			&+\quad J_{xy} (\cos k_x - \cos k_y)\, \Gamma_5 +J_{xz} (\cos k_z - \cos k_x)\, \Gamma_5\ ,
		\end{aligned}
		\label{eq:Hamiltonian}
	\end{equation}
	where, \(\lambda\), \(t\), and \(m_0\) denote the spin-orbit coupling (SOC) strength, nearest-neighbour hopping amplitude, and staggered mass term with \(J_{xy}\), \(J_{xz}\) indicate the coupling strengths of the AMs with $d_{x^2-y^2}$ and $d_{x^2-z^2}$ symmetry respectively. The $\Gamma$ matrices are defined as
	$\Gamma_1=\mu_x\sigma_x s_y$, $\Gamma_2=\mu_z\sigma_0 s_0$, $\Gamma_3=\mu_x\sigma_z s_y$,
	$\Gamma_4=\mu_x\sigma_y s_y$, $\Gamma_5=\mu_x\sigma_0 s_z$
	where, $\mu_i$, $\sigma_i$, and $s_i$ are Pauli matrices acting on sublattice, orbital, and spin spaces. 
	Note that, our chosen AM exchange terms satisfy the defining symmetries of $d$-wave AMs. Specifically both the terms, $\mc{H}_{xy}^{\rm AM}(\mbf{k})$ and $\mc{H}_{xz}^{\rm AM}(\mbf{k})$, breaks time reversal symmetry (TRS) i.e., $\mathcal{T} \mc{H}^{\rm AM}_{xy}(\mbf{k}) \mathcal{T}^{-1} \ne \mc{H}^{\rm AM}_{xy}(-\mbf{k})$ and $\mathcal{T} \mc{H}^{\rm AM}_{xz}(\mbf{k}) \mathcal{T}^{-1} \ne \mc{H}^{\rm AM}_{xz}(-\mbf{k})$ where $\mc{T}=i \mu_0 \sigma_0 s_y \mathcal{K}$ is the TRS operator. In addition, both these terms also preserves the required $C_4\mc{T}$-symmetry for $d$-wave AMs. Particularly, $d_{x^2-y^2}$ AM term remains invariant under $C_{4z}\mc{T}$ transformation whereas $d_{x^2-z^2}$ AM under $C_{4y}\mc{T}$ transformation which can be shown in the following way. Under a $C_{4z}$ rotation, $(k_x,k_y,k_z)$ transforms as $(k_x, k_y,k_z) \rightarrow (k_y, -k_x,k_z)$ which leads to,  
	\begin{equation*}
		\cos k_x - \cos k_y \rightarrow \cos k_y - \cos k_x = -(\cos k_x - \cos k_y)\ .
	\end{equation*}
	Moreover, under TRS operation, $\mathcal{T} \mc{H}^{\rm AM}_{xy}(\mbf{k}) \mathcal{T}^{-1} = - \mc{H}^{\rm AM}_{xy}(\mbf{k})$. Therefore, under combined action of $C_{4z}\mc{T}$, $\mc{H}^{\rm AM}_{xy}(\mbf{k})$ does not change i.e. $(C_{4z}\mc{T}) \mc{H}^{\rm AM}_{xy}(\mbf{k}) (C_{4z}\mc{T})^{-1} = \mc{H}^{\rm AM}_{xy}(\mbf{k})$.
	
	Following the similar lines of arguments, one can show that $\mc{H}^{\rm AM}_{xz}(\mbf{k})$ remains invariant under $C_{4y}\mc{T}$ symmetry where $(k_x,k_y,k_z)$ transforms as $(k_x, k_y,k_z) \rightarrow (-k_z, k_y,k_x)$  under $C_{4y}$ transformation. Therefore, our chosen terms for exchange couplings satisfies the defining relations (symmetry requirements) of a $d$-wave AM.

	%
	%
	
	
	\vspace{0.2cm}
    $\bullet$	\underline{\textbf{Slab geometry:}}
	In a three dimensional (3D) system, slab geometry corresponds to the situation when 
	the system has finite size along any one direction, say $z$-direction, while the other two directions, $x$ and $y$-directions, are considered to be in the momentum space so that $k_x,k_y$ remain good quantum numbers. Usually, open boundary condition (OBC) is applied along the finite size direction. This geometry is utilized to understand the nature of the surface states present in the system. Therefore, considering slab geometry, the Hamiltonian can be written as:
	\begin{equation}
		\begin{aligned}
			H(k_x,k_y,z) = \sum_{z} c_{k_x,k_y,z}^{\dagger} \bigg[ & \bigg(\left\{ m_0-6t+2t(\cos k_x+\cos k_y)\right\} \Gamma_4+2\lambda (\sin k_x \Gamma_1+\sin k_y \Gamma_2)+J_{xy}(\cos k_x-\cos k_y)\Gamma_5\\
			&-J_{xz}\cos k_x \Gamma_5\bigg) c_{k_x,k_y,z}+ \left(t\Gamma_4-i\lambda_z\Gamma_3+\frac{J_{xz}}{2}\right)c_{k_x,k_y,z+1}
			\bigg] + h.c.\ ,
			\label{eq:Hamiltonianslab}
		\end{aligned}
	\end{equation}
	

	$\bullet$ \underline{\textbf{Rod geometry:}} 
	\vspace{0.2cm}
	In rod geometry, the system exhibits finite size along any two directions while the other direction is considered to be in the momentum space. The spectral properties in the rod geometry is utilized to understand the nature of the hinge localized modes which are the manifestation of second-order topological insulator (SOTI) phase. 
	In the main text, we have used two types of rod geometry calculations considering finite size along (i) $x$ and $y$ directions, and (ii) $x$ and $z$ directions. 
	Considering such geometry, the Hamiltonian with finite size along $x$ and $y$ directions can be written as,
	\begin{equation}
		\begin{aligned}
			H(x,y,k_z) = \sum_{x,y} c_{x,y,k_z}^{\dagger} \bigg[ & \bigg(\left\{ m_0-6t+2t(\cos k_z)\right\} \Gamma_4+2\lambda (\sin k_z) \Gamma_3+J_{xz}\cos k_z \Gamma_5\bigg) c_{x,y,k_z}\\ &+\left(t\Gamma_4-i\lambda_x\Gamma_1-\frac{J_{xz}}{2}+\frac{J_{xy}}{2}\right)c_{x+1,y,k_z}+\left(t\Gamma_4-i\lambda_y\Gamma_2-\frac{J_{xy}}{2}\right)c_{x,y+1,k_z}
			\bigg] + h.c.\ ,
			\label{eq:Hamiltonianrod_xy}
		\end{aligned}
	\end{equation}
	
	Similarly, Hamiltonian with finite size along $x$ and $z$ can be written as,
	\begin{equation}
		\begin{aligned}
			H(x,k_y,z) = \sum_{x,z} c_{x,k_y,z}^{\dagger} \bigg[ & \bigg(\left\{ m_0-6t+2t(\cos k_y)\right\} \Gamma_4+2\lambda (\sin k_y) \Gamma_2-J_{xy}\cos k_y \Gamma_5\bigg) c_{x,k_y,z}\\ &+\left(t\Gamma_4-i\lambda_x\Gamma_1-\frac{J_{xz}}{2}+\frac{J_{xy}}{2}\right)c_{x+1,k_y,z}+\left(t\Gamma_4-i\lambda_z\Gamma_3+\frac{J_{xz}}{2}\Gamma_5\right)c_{x,k_y,z+1}
			\bigg] + h.c.\ ,
			\label{eq:Hamiltonianrod_xz}
		\end{aligned}
	\end{equation}
	
	$\bullet$ \underline{\textbf{Cubic geometry:}} 
	\vspace{0.2cm}
	In the main text, we have used the cubic geometry to illustrate the local density of states (LDOS). In such geomery, the system has finite size along all the three directions and the corresponding Hamiltonian can be written as, 
	\begin{equation}
		\begin{aligned}
			H = \sum_{i,j,k} c_{i,j,k}^{\dagger} \bigg[ & \left\{m_0-6t\right\} \Gamma_4 c_{i,j,k} 
			+\left(t \Gamma_4-i\lambda_x\Gamma_1+\frac{J_{xy}-J_{xz}}{2}\Gamma_5 \right)c_{i+1,j,k} \\
			& +\left(t\Gamma_4-i\lambda_y\Gamma_2-\frac{J_{xy}}{2}\right)c_{i,j+1,k}+\left(t\Gamma_4-i\lambda_y\Gamma_3+\frac{J_{xz}}{2}\right)c_{i,j,k+1} \bigg] + h.c.\ ,
			\label{eq:HamiltonianRealspace}
		\end{aligned}
	\end{equation}
	where, \[
	c_{i,j,k} =
	\big(
	c_{i,j,k}^{A,\alpha,\uparrow},
	c_{i,j,k}^{A,\alpha,\downarrow},
	c_{i,j,k}^{A,\beta,\uparrow},
	c_{i,j,k}^{A,\beta,\downarrow},
	c_{i,j,k}^{B,\alpha,\uparrow},
	c_{i,j,k}^{B,\alpha,\downarrow},
	c_{i,j,k}^{B,\beta,\uparrow},
	c_{i,j,k}^{B,\beta,\downarrow}
	\big)^{T}\ .
	\]
	
	Similarly, in Eq.~(\ref{eq:Hamiltonianslab}), Eq.~(\ref{eq:Hamiltonianrod_xy}), 
	and Eq.~(\ref{eq:Hamiltonianrod_xz})
	\[
	c_{k_x,k_y,z} =
	\big(
	c_{z}^{A,\alpha,\uparrow},
	c_{z}^{A,\alpha,\downarrow},
	c_{z}^{A,\beta,\uparrow},
	c_{z}^{A,\beta,\downarrow},
	c_{z}^{B,\alpha,\uparrow},
	c_{z}^{B,\alpha,\downarrow},
	c_{z}^{B,\beta,\uparrow},
	c_{z}^{B,\beta,\downarrow}
	\big)_{k_x,k_y}^{T}\ ,
	\]
	
	\[
	c_{x,y,k_z} =
	\big(
	c_{x,y}^{A,\alpha,\uparrow},
	c_{x,y}^{A,\alpha,\downarrow},
	c_{x,y}^{A,\beta,\uparrow},
	c_{x,y}^{A,\beta,\downarrow},
	c_{x,y}^{B,\alpha,\uparrow},
	c_{x,y}^{B,\alpha,\downarrow},
	c_{x,y}^{B,\beta,\uparrow},
	c_{x,y}^{B,\beta,\downarrow}
	\big)_{k_z}^{T}\ ,
	\]
	
	\[
	c_{x,k_y,z} =
	\big(
	c_{x,z}^{A,\alpha,\uparrow},
	c_{x,z}^{A,\alpha,\downarrow},
	c_{x,z}^{A,\beta,\uparrow},
	c_{x,z}^{A,\beta,\downarrow},
	c_{x,z}^{B,\alpha,\uparrow},
	c_{x,z}^{B,\alpha,\downarrow},
	c_{x,z}^{B,\beta,\uparrow},
	c_{x,z}^{B,\beta,\downarrow}
	\big)_{k_y}^{T}\ ,
	\]
	
	Here, $A$ and $B$ denote sublattice index and $\alpha,\beta$ are orbital index.
	
		\begin{center}
		\specialsubsection{\hskip 0.4cm Hybrid-order topological  phase} 
	\end{center}
	Here we use Eq.~(\ref{eq:Hamiltonian}) and present the surface as well as hinge spectrum by considering slab and rod geometry respectively. To construct a slab, we impose OBC along one direction while maintaining periodic boundary conditions (PBC) along the other two. In Fig.~\ref{FigS1}(a) and (e), we apply OBC along the $x$ direction and PBC along $y$ and $z$, thereby probing the $yz$ plane. Considering the limit $k_y=0$, we find that the surfaces along $x$ remain gapless in case of the 3D TI. However, once the two-dimensional altermagnet term is introduced with $J_{xy}=2t$, these surfaces become gapped. A similar behavior is observed in Fig.~\ref{FigS1}(b) and (f), where we study the $y$ surfaces by imposing OBC along $y$, PBC along $x$ and $z$, and fixing $k_z=0$. In contrast, when we examine the $z$ surfaces in Fig.~\ref{FigS1}(c) and (g) with OBC along $z$ and PBC along $x$ and $y$, we observe that the surfaces remain gapless even after turning on $J_{xy}$. Also in the first case, we observe that the zero-energy LDOS gives rise to modes along all surfaces as illustrated in Fig.~\ref{FigS1}(d). However, in contrast we find the surfaces are gapped for $x$ and $y$ appearing hinges, but the surfaces along $z$ remains gapless as depicted in Fig.~\ref{FigS1}(h).
	\begin{figure}[h]
		\centering
		\includegraphics[width=1.0\textwidth]{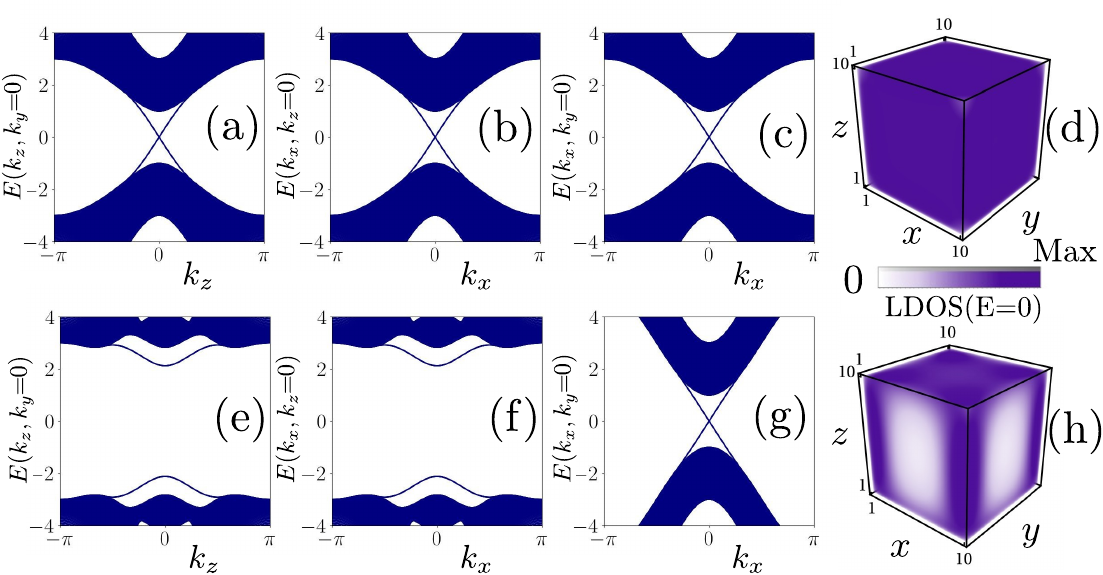}
		\caption{2D Surface states and corresponding LDOS for different values of $J_{xy}$ are shown. Panels (a–c) exhibit the surface spectra for $J_{xy}=0$, with the corresponding LDOS at $E=0$ presented in panel (d). Panels (e–g) display the surface spectra for $J_{xy}=2t$, and the corresponding LDOS is shown in panel (h). Other model parameters are set to be $m_0=t$, $\lambda=t$ and $t=1$.
		}
		\label{FigS1}
	\end{figure}
	\begin{center}
			\specialsubsection{\hskip 0.4cm SOTI phase} 
	\end{center}

	Here, we analyze the band spectrum of our system in the rod geometry considering finite $J_{xz}$ along with $J_{xy}$. We explore the appearance of hinge states employing 
	OBC along two directions and PBC along the remaining direction. In Fig.~\ref{FigS2}(a) we show that the surface normal to the $z$-direction becomes gapped after introducing the finite coupling $J_{xz}$ in addition to $J_{xy}$. For $J_{xz}<J_{xy}$ (SOTI$^{\text{(I)}}$), the spectrum with PBC along $z$ and OBC along $x,y$ directions exhibits gapless 1D hinge modes propagating along $z$-direction (see Fig.~\ref{FigS2}(b)). 
	On the other hand, for $J_{xz}>J_{xy}$ (SOTI$^{\text{(II)}}$), those hinge modes become gapped as shown in Fig.~\ref{FigS2}(c).
	\begin{figure}[h]
		\centering
		\includegraphics[width=1.0\textwidth]{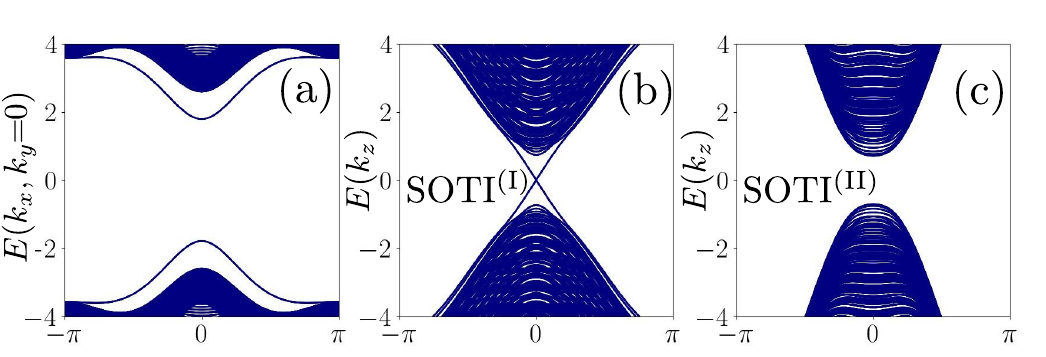}
		\caption{We show the 2D surface state in panel (a), calculated employinmg OBC along the $x$ and $y$ directions, and with fixed momentum at $k_{y} = 0$. In panels (b) and (c), we display the hinge states in the SOTI phase of type I and type II, respectively. For panels (a) and (b), the model parameters are chosen as $J_{xy} = 4t$ and $J_{xz} = 2t$ respectively, while for panel (c) the same model parameters are set to $J_{xy} = 2t$ and $J_{xz} = 4t$.}
		\label{FigS2}
	\end{figure}
	
			\begin{center}
		\specialsection{\hskip 0.4cm Topological Invariant}\label{Sec:III}
	\end{center}

	We here provide the detailed formulation of the winding number ($W$) and the quadrupolar winding number ($N$).  
	\begin{center}
			\specialsubsection{\hskip 0.4cm Winding Number $W$}
	\end{center}

	We start from the Hamiltonian presented in Eq.~\eqref{eq:Hamiltonian} and set $k_x = k_y = 0$ and $J_{xz}=0$. Then Eq.~\eqref{eq:Hamiltonian} reduces to  
	\begin{equation}
		\mathcal{H}= 2\lambda \sin k_z \Gamma_3
		+ \left[ m_0 - 2t + 2t \cos k_z \right] \Gamma_4 \ .
		\label{eq:Hamiltonian_w}
	\end{equation}  
	
	We introduce the chiral operator as $U_s = \mu_0\sigma_x s_0$. In this chiral basis, Eq.~\eqref{eq:Hamiltonian_w} takes an off-diagonal form:  
	\[
	\mathcal{H} = U_c \mathcal{H} U_c^\dagger =
	\begin{bmatrix}
		0 & h \\
		h^\dagger & 0 \ 
	\end{bmatrix}\ ,
	\]  
	where, $U_c$ is a unitary matrix constructed from the eigenvectors of the chiral basis and $h$ has the form $h(k)=[-2i (m_0 - 2 t + 2 t \cos k_z) + 4  \lambda  \sin k_z]\sigma_xs_y$.
	
	The corresponding winding number is defined as \cite{Mondal_2025,Chiu_2016} 
	\begin{equation}
		W=\frac{i}{2\pi}\int_{\text{BZ}} dk \ \text{Tr}\!\left[ h^{-1}(k)\,\partial_k h(k) \right]\ .
	\end{equation}  
	
	\begin{center}
	\specialsubsection{\hskip 0.4cm Quadrupolar Winding Number $N_{xy}$}
\end{center}

	To compute the quadrupolar winding number in phase-1, we set $k_z = 0$ and impose PBC along $x$ and $y$.  
	
	The Hamiltonian preserves chiral symmetry, satisfying  
	\[
	S^{\dagger} H S = -H\ ,
	\]  
	with the chiral operator chosen as $S = \mu_y \sigma_z s_z$. In this basis, the Hamiltonian again reduces to a block off-diagonal form:  
	\[
	H = U_c H U_c^\dagger =
	\begin{bmatrix}
		0 & h \\
		h^\dagger & 0
	\end{bmatrix}\ .
	\]  
	
	Here, the system is partitioned into two sublattice degrees of freedom, labeled by $A$ and $B$, corresponding to the eigenvalues $+1$ and $-1$ of $S$. 
	Thus, the eigenstates of $H$ can be expressed as  
	\[
	\ket{\psi_n} =
	\begin{bmatrix}
		\ket{\psi_n^A} \\
		\ket{\psi_n^B}
	\end{bmatrix}\ ,
	\]  
	with $\ket{\psi_n^A}$ and $\ket{\psi_n^B}$ denoting the normalized components in the respective subspaces.  
	
	Performing a singular value decomposition (SVD) of the off-diagonal block yields  
	\[
	h = U_A \Sigma U_B^\dagger \ ,
	\]  
	where, $U_A$ and $U_B$ contain the singular vectors, and $\Sigma$ is diagonal with singular values.  
	
	The quadrupole operator is defined as~\cite{Benalcazar_2022}  
	\begin{equation}
		Q = \exp\left( -\frac{i 2\pi x y}{L_x L_y} \right) \ ,
	\end{equation}  
	and the corresponding sublattice-resolved quadrupole moments are given by  
	\begin{equation}
		Q_{A,B} = \sum_{R, u \in A, B} \ket{R, u}\, Q \,\bra{R, u}.
	\end{equation}  
	
	Projecting these onto the SVD eigenbasis yields  
	\begin{equation}
		\bar{Q}_{A,B} = U_{A,B}^\dagger Q U_{A,B}\   .
	\end{equation}  
	
	Finally, the {\bf{quadrupolar winding number}} can be expressed as~\cite{Benalcazar_2022,Pal_2025,Subhadarshini_2025}:  
	\begin{equation}
		N_{xy} = \frac{1}{2\pi i}\, \operatorname{Tr}\, \log \!\left( \bar{Q}_{x,y}^A \, \bar{Q}_{x,y}^{B\dagger} \right)\ .
	\end{equation} 
	In phase-2, one can analogously compute:  
	\begin{itemize}
		\item $N_{xy}$ in the SOTI$^{(\text{I})}$ phase (considering OBC along $x,y$, $k_z=0$), and  
		\item $N_{xz}$ in the SOTI$^{(\text{II})}$ phase (considering OBC along $x,z$, $k_y=0$),  
	\end{itemize}  
	employing the chiral operator for $N_{xy}$ and $N_{xz}$ as $U = \mu_y \sigma_z s_z$ and $U = \mu_0 \sigma_0 s_x$ respectively.  
	
	\begin{center}
		\specialsection{\hskip 0.4cm Surface theory}\label{Sec:IV}
	\end{center}
	We rewrite the tight-binding Hamiltonian described in Eq.~\eqref{eq:Hamiltonian} as a low energy contunuum model by replacing $\sin k_i \sim k_i$ and $\cos k_i \sim (1-k_i^2/2)$ with $i=x,y,z$. This is given by,
	\begin{equation}
		H = 2\lambda \sum_{j=1}^3 k_j \Gamma_j 
		+ \left(m_0 - t \sum_{j=1}^3 k_j^2\right)\Gamma_4
		- \frac{J_{xy}}{2}\left(k_x^2 - k_y^2\right)\Gamma_5
		- \frac{J_{xz}}{2}\left(k_z^2 - k_x^2\right)\Gamma_5 \ .
		\label{eq:Low_energy_Ham}
	\end{equation}
	Here, we perform the analytical calculation regarding mass inversion exhibiting surface and hinge theory. We use Eq.~(\ref{eq:Low_energy_Ham}) to derive surface and hinge Hamiltonian as we discuss later.
	
	\vspace{0.2cm}
	
	\underline{\textbf{$xy$ surface}}:-
	\vspace{0.1cm}
	
	To derive the surface Hamiltonian on the $xy$ surface, we impose OBC along the $z$-direction and PBC along $x$ and $y$ directions. Substituting $k_z \rightarrow -i\partial_z$ and retaining only the linear terms in $k_x$ and $k_y$, the Hamiltonian in Eq.~(\ref{eq:Low_energy_Ham}) separates into two parts \cite{Arnob_2021,Subhadarshini_2025}:
	\begin{equation}
		\begin{aligned}
			H_I &= (m_0 + t\partial_z^2)\Gamma_4 - 2i\lambda\partial_z \Gamma_3\ , \\
			H_{II} &= 2\lambda k_x \Gamma_1 + 2\lambda k_y \Gamma_2 + \frac{J_{xz}}{2}\partial_z^2 \Gamma_5 \ .
		\end{aligned}
	\end{equation}
	
	We first solve the eigenvalue problem $H_I |\psi\rangle = 0$ with the boundary condition $|\psi(z)\rangle \to 0$ as $z \to 0, \infty$. The solution takes 
	the form
	\begin{equation}
		|\psi\rangle = A e^{-k_1 z}\sin(k_2 z)\, e^{i(k_x x + k_y y)} |\chi\rangle \ ,
	\end{equation}
	where,
	\begin{equation}
		k_1 = \frac{\lambda}{t}, 
		\quad k_2 = \sqrt{\frac{m_0}{t} - \frac{\lambda^2}{t^2}}, 
		\quad |A|^2 = \frac{4k_1(k_1^2 + k_2^2)}{k_2^2} \ .
	\end{equation}
	
	Here, $|\chi\rangle$ is an 8-component spinor, which can be chosen as
	\begin{equation}
		\chi_{1}=
		\begin{pmatrix}
			0\\0\\0\\0\\0\\1\\0\\1
		\end{pmatrix}, \quad 
		\chi_{2}=
		\begin{pmatrix}
			0\\0\\0\\0\\1\\0\\1\\0
		\end{pmatrix}, \quad 
		\chi_{3}=
		\begin{pmatrix}
			0\\1\\0\\1\\0\\0\\0\\0
		\end{pmatrix}, \quad 
		\chi_{4}=
		\begin{pmatrix}
			1\\0\\1\\0\\0\\0\\0\\0
		\end{pmatrix}.
	\end{equation}
	
	The effective surface Hamiltonian matrix elements are obtained as
	\begin{equation}
		H^{S}_{xy,\alpha\beta} = \int_0^\infty dz \, \langle \psi_\alpha | H_{II} | \psi_\beta \rangle \ ,
	\end{equation}
	with $\alpha,\beta = 1,\dots,8$. This yields the $xy$ surface Hamiltonian as
	\begin{equation}
		H_{xy}^S = -2\lambda k_x \sigma_x s_y - 2\lambda k_y \sigma_z s_0 + \frac{J_{xz} m_0}{2t}\sigma_x s_z \ .
	\end{equation}
	
	{\underline{\textbf{$yz$ surface}}}:-
	
	\vspace{0.1cm}
	For the $yz$ surface, we impose OBC along $x$, while $y$ and $z$ remain periodic. Thus the Hamiltonian becomes
	\begin{equation}
		\begin{aligned}
			H_I &= (m_0+t\partial_x^2)\Gamma_4 - 2i\lambda \partial_x \Gamma_1 \ ,\\
			H_{II} &= 2\lambda k_y \Gamma_2 + 2\lambda k_z \Gamma_3 
			+ \frac{J_{xy}-J_{xz}}{2}\partial_x^2 \Gamma_5 \ .
		\end{aligned}
	\end{equation}
	
	With the same approximation, the corresponding zero-energy mode takes the form
	\begin{equation}
		|\psi\rangle = A e^{-k_1 x}\sin(k_2 x)\, e^{i(k_yy+k_zz)}|\Xi\rangle \ ,
	\end{equation}
	where, the four independent spinors are
	\begin{equation}
		\Xi_{1}=\begin{pmatrix}0\\0\\0\\0\\0\\0\\0\\1\end{pmatrix}, \quad
		\Xi_{2}=\begin{pmatrix}0\\0\\0\\0\\0\\0\\1\\0\end{pmatrix}, \quad
		\Xi_{3}=\begin{pmatrix}0\\0\\0\\1\\0\\0\\0\\0\end{pmatrix}, \quad
		\Xi_{4}=\begin{pmatrix}0\\0\\1\\0\\0\\0\\0\\0\end{pmatrix}.
	\end{equation}
	
	Thus the projected surface Hamiltonian can be written as
	\begin{equation}
		H_{yz}^S = -2\lambda k_y \sigma_z s_0 + 2\lambda k_z \sigma_x s_y 
		+ \frac{(J_{xy}-J_{xz})m_0}{2t}\,\sigma_x s_z \ .
		\label{eq:Eq_yz_s}
	\end{equation}
	
	{\underline{\textbf{$xz$ surface}}}:-
	\vspace{0.1cm}
	
	In case of $xz$ surface, we impose OBC along $y$ direction while maintaining PBC along $x$ and $z$ directions. 
	The Hamiltonian reads
	\begin{equation}
		\begin{aligned}
			H_I &= (m_0+t\partial_y^2)\Gamma_4 - 2i\lambda \partial_y \Gamma_2 \ ,\\
			H_{II} &= 2\lambda k_x \Gamma_1 + 2\lambda k_z \Gamma_3 
			-\frac{J_{xy}}{2}\partial_y^2 \Gamma_5 \ .
		\end{aligned}
	\end{equation}
	
	The zero-energy state is
	\begin{equation}
		|\psi\rangle = A e^{-k_1 y}\sin(k_2 y)\, e^{i(k_xx+k_zz)}|\xi\rangle \ ,
	\end{equation}
	with four independent spinors as
	\begin{equation}
		\xi_{1}=\begin{pmatrix}-i\\0\\0\\0\\0\\0\\0\\1\end{pmatrix}, \quad
		\xi_{2}=\begin{pmatrix}0\\i\\0\\0\\0\\0\\1\\0\end{pmatrix}, \quad
		\xi_{3}=\begin{pmatrix}0\\0\\i\\0\\0\\1\\0\\0\end{pmatrix}, \quad
		\xi_{4}=\begin{pmatrix}0\\0\\0\\-i\\1\\0\\0\\0\end{pmatrix}.
	\end{equation}
	
	The corresponding surface Hamiltonian becomes
	\begin{equation}
		H_{xz}^S = -2\lambda k_x \sigma_z s_0 + 2\lambda k_z \sigma_x s_0 
		+ \frac{J_{xy}m_0}{2t}\, \sigma_y s_x \ .
		\label{eq:Eq_xz_s}
	\end{equation}
	
	From Eq.~\eqref{eq:Eq_yz_s} and Eq.~\eqref{eq:Eq_xz_s}, we observe that at $k_x=k_y=k_z=0$ both $yz$ and $xz$ surfaces are gapped. Under unitary transformations
	\[
	u_1 = \sigma_0 e^{-i\frac{\pi}{4} s_y}, 
	\qquad 
	u_2 = \sigma_z e^{i\frac{\pi}{4} s_0},
	\]
	with $u_2^{-1}u_1^{-1} H_{xz}u_1u_2$, the surface Hamiltonians simplify to:
	\begin{equation}
		\begin{aligned}
			H_{xy}^S &= \frac{J_{xz} m_0}{2t}\sigma_x s_z \ , \\
			H_{yz}^S &= \frac{(J_{xy} - J_{xz}) m_0}{2t}\sigma_x s_z\ ,\\
			H_{xz}^S &= -\frac{J_{xy} m_0}{2t}\sigma_x s_z \ .
		\end{aligned}
		\label{eq:Eq_phase_2_s}
	\end{equation}
	
	From Eq.~\eqref{eq:Eq_phase_2_s}, it is evident that the surface Dirac masses exhibit relative sign changes. In phase-1 (HyOTP), $J_{xz}=0$, hence the surface along $xy$ is gapless and mass changes between $xz$ and $yz$ surfaces. According to Jackiew-Rebbi theorem~\cite{Jackiw_1976}, we have 1D hinge modes propagating along the $z$-direction. 
	
	In phase-2 (SOTI), from Eq.~\eqref{eq:Eq_phase_2_s}, between the $xy$ and $xz$ surfaces, the mass terms always have opposite signs resulting in $x$-directed hinge modes. On top of that, if $J_{xz} > J_{xy}$, an additional sign change occurs between the mass terms of $xy$ and $yz$ surfaces, leading to hinge modes along the $x$ and $y$ directions (SOTI$^{(\text{II})}$). Conversely, if $J_{xy} > J_{xz}$, the additional sign change appears between the $yz$ and $xz$ surfaces, giving rise to hinge modes along the $x$ and $z$ directions (SOTI$^{(\text{I})}$).

	\begin{center}
	\specialsection{\hskip 0.4cm Details of transport calculation}\label{Sec:V}
\end{center}		
	
	In the main text, we have computed the two terminal differential conductance, $dI/dV$, to identify the surface states as well as hinge states in different topological phases. Here, we discuss the computational details of the transport calculation. To obtain the differential conductance, we first construct the scattering matrix, which relates the incoming propagating modes to the outgoing modes in the leads with the central system being considered 
	as the scatterer. Incoming, outgoing states and the scattering matrix are defined as:
	\begin{eqnarray}
		\Psi^{\rm in}=& [\psi_1^{\rm L},\psi_2^{\rm L},...,\psi_{N_L}^{\rm L},\psi_1^{\rm R},\psi_2^{\rm R},...,\psi_{N_R}^{\rm R}]^T\ , \\ 
		\Phi^{\rm out}=&[\phi_1^{\rm L},\phi_2^{\rm L},...,\phi_{N_L}^{\rm L},\phi_1^{\rm R},\phi_2^{\rm R},...,\phi_{N_R}^{\rm R}]^T \ ,
	\end{eqnarray}
	\begin{center}
		$\Phi^{\rm out} = \hat{S}\Psi^{\rm in}$\ ,
	\end{center}
	where, $\psi^{\rm L(R)}_i$ is the incoming state from the left (right) lead in the $i^{\rm{th}}$ mode. Here, $N_L(N_R)$ is the number of occupied modes/channels in the left (right) lead for a given voltage bias $eV$. Similarly, $\phi^{\rm L(R)}_i$ is the outgoing state into the left (right) lead in the $i^{\rm{th}}$ mode after the scattering event takes place. Here, `$T$' denotes the tranpose operation. 
	The unitary scattering matrix $\hat{S}$ of dimension $(N_L+N_R)\times(N_L+N_R)$ reads
	\begin{equation}
		\hat{S}=
		\begin{bmatrix}
			\hat{r} & \hat{t^\prime} \\
			\hat{t} & \hat{r^\prime} \ 
		\end{bmatrix}\ ,
	\end{equation}
	where, $\hat{r} \,(\hat{r^\prime})$ is a square matrix of dimension $N_L(N_R)$ and $\hat{t}\, (\hat{t^\prime})$ is a matrix of dimension $N_R\!\!\times\!\! N_L(N_L\!\!\times\! \!N_R)$. Physically, $\hat{r}\,(\hat{r^\prime})$ represents the reflection matrix with elements $r_{i,j}\,(r^\prime_{i,j})$ being the amplitude of reflection from the $j^{\rm{th}}$ mode to the $i^{\rm{th}}$-mode in the left (right) lead. Similarly, $\hat{t}\,(\hat{t^\prime})$ denotes the transmission matrix with elements $t_{i,j}\,(t^\prime_{i,j})$ denoting the amplitude of the transmission from the $j^{\rm{th}}$ mode in left (right) lead 
	to the $i^{\rm{th}}$ mode in the right (left) lead following the unitarity condition: $r^\dagger r + t^\dagger t = \mc{I}$. Within this formalism, the two-terminal differential conductance at zero temperature can be obtained using the Landauer formula given by~\cite{datta_1995},
	\begin{equation}
		\frac{dI}{dV} \,(eV)=G_0 \, \text{Tr}[t^\dagger(E) t(E)]|_{E=eV}\ , 
		\label{Eq.Landauer Formula}
	\end{equation}
	where $G_0=e^2/h$ is the unit of quantum conductance. The scattering amplitudes can be calculated numerically using python package KWANT~\cite{Groth_2014}.
	
	For the surface states in the hybrid order topological phase, we attach two leads at $x=1$ and $x=L_x$ (see inset of Fig.\,1(a) of the main text) and apply a voltage bias $eV$ across the leads. Importantly, we model both the leads and the system using the Hamiltonian in the slab geometry as mentioned in Eq.\,\eqref{eq:Hamiltonianslab}. The leads are attached along the $x$-direction which capture all the transverse modes propagating along the $x$-direction. However, the surface states also have propagating modes along the $y$-direction. To incorporate these modes, we sum over all $k_y$ in the $dI/dV$ calculation as,
	
	\begin{equation}
		\frac{dI^{\rm (S)}}{dV} \,(eV)=G_0 \, \sum_{k_y}\text{Tr}[t_S^\dagger(E) t_S(E)]|_{E=eV}\ , 
		\label{Eq.Landauer Formula_slab}
	\end{equation}
	where, $t_S(E)$ is the scattering the matrix in the transport setup mentioned above. 
	
	Similarly, for the hinge states which propagate along the $z$-direction, we attach two leads at $z=1$ and $z=L_z$, and apply a voltage bias $eV$ across the leads. Importantly, for this case, we model the leads and the system using the Hamiltonian in the rod geometry as mentioned in  Eq.\,\eqref{eq:Hamiltonianrod_xy}. Since in this case, the system carries propagating modes only along the $z$-direction while the modes are localized in the other directions, we use the following expression to obtain the $dI/dV$, 
	\begin{equation}
		\frac{dI^{\rm (H)}}{dV} \,(eV)=\frac{e^2}{h} \, \text{Tr}[t_H^\dagger(E) t_H(E)]|_{E=eV}\ , 
		\label{Eq.Landauer Formula_hinge}
	\end{equation}
	where, $t_H(E)$ denotes the scattering matrix containing all the information about the transverse modes in this geometry.

		\begin{center}
		\specialsection{\hskip 0.4cm Exponentially decaying profile of altermagnetic exchange}\label{Sec:VI}
	\end{center}
	\begin{figure}[h!]
		\centering
		\includegraphics[width=0.9\textwidth]{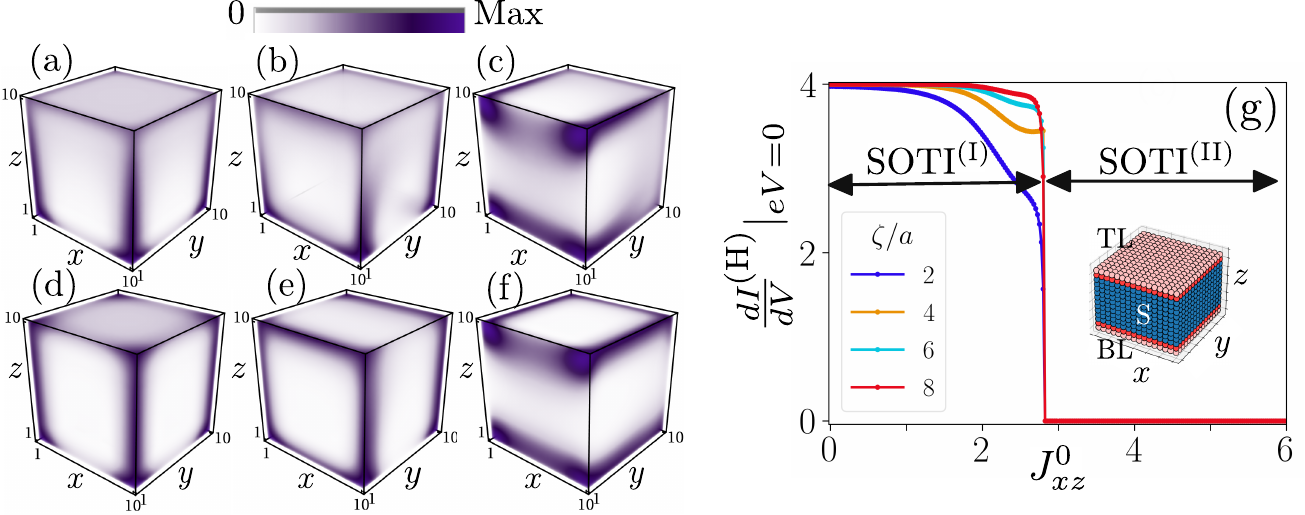}
		\caption{We depict the LDOS at $E = 0$ for different values of the decaying altermagnetic strength (localization lengths). Panels (a), (b), and (c) correspond to $\zeta = 4a$, while panels (d), (e), and (f) refer to $\zeta = 9a$. In panels (a) and (d), we set $J_{xz}^0 = 0$ and $J_{xy}^0 = 3t$. Panels (b) and (e) correspond to $J_{xz}^0 = t$ and $J_{xy}^0 = 3t$, whereas panels (c) and (f) correspond to $J_{xz}^0 = 3t$ and $J_{xy}^0 = t$. The remaining model parameters are fixed to $m_0 = 2t$, $\lambda = t$, and $t = 1$. For  panels (a)-(f), the calculations are performed on a $10 \times 10 \times 10$ finite lattice size. (g) $\frac{dI^{\rm (H)}}{dV}(eV=0)$ is depicted as a function of $J_{xz}^0$ for various strengths of $\zeta/a$ with model parameters chosen as, $(m_0,\lambda,J_{xy}^{0})=(t,t,3t)$ and $(L_x,L_y,L_z)=(20,20,10)$.
		}
		\label{FigS3}
	\end{figure}
	In this section, we demonstrate that the essential physics qualitatively survives even when the altermagnetic couplings $J_{xy}$ and $J_{xz}$ are not uniform throughout the bulk of TI, instead decay over a finite number of layers. Specifically, we assume a physically reasonable decay profile of the altermagnetic exchange terms that decays exponentially into the bulk as, $J_{xy}(z)=J_{xy}^0 e^{-z/\zeta}$ and $J_{xz}(y)=J_{xz}^0 e^{-y/\zeta}$. Here, $J_{xy}^0$ and $J_{xz}^0$ denote the strengths of altermagnetic exchange terms at the two interfaces (see Fig.~1(b) of the main text for the setup), and $\zeta$ is the localization length of the altermagnetic exchange terms (assumed symmetric) along both the $z$ and $y$-directions \ie normal to the respective interfaces. 
	Thus, Eq.~(3) and (4) of the main text can be rewritten as, 
	\begin{eqnarray}
		\mc{H}^{\rm AM}_{xy}(\mbf{k}) &=& J_{xy}(z) (\cos k_x - \cos k_y)\, \Gamma_5\ , \\
		\mc{H}^{\rm AM}_{xz}(\mbf{k}) &=& J_{xz}(y) (\cos k_z - \cos k_x)\, \Gamma_5\ .
	\end{eqnarray}
	With these modifications, we first examine the stability of the hybrid-order topological phase (HyOTP), SOTI$^{\rm (I)}$, and SOTI$^{\rm (II)}$ by computing local density of states (LDOS) at zero energy.

	Variation of LDOS $(E=0)$ for the HyOTP is depicted in Fig.\,\ref{FigS3}(a) and (d) considering $\zeta=4a$ and $\zeta=9a$, respectively ($a$ denote the lattice spacing, set to unity). For $\zeta = 4a$, the hinge modes decay rapidly along the $z$ direction, resulting in a higher density of states at the bottom surface compared to the top surface. This leads to a pronounced asymmetry in the LDOS at $E=0$. In contrast, for $\zeta = 9a$, the decay extends over more layers, and the hinge modes become nearly symmetric across the top and bottom surfaces. On the other hand, Figs.~\ref{FigS3}(b) and (e) correspond to the variation of LDOS $(E=0)$ in the SOTI$^{\text{I}}$ phase for $\zeta = 4a$ and $\zeta = 9a$, respectively. In both the cases, the surface states on the bottom surface are fully gapped, and no hinge modes appear propagating along the $y$ direction. On the top surface, the hinge modes exhibit a slight dispersion for $\zeta = 4a$, while for $\zeta = 9a$ [Fig.~\ref{FigS3}(e)] the hinge modes along the $xz$ direction become clearly visible. Moreover, Figs.\,\ref{FigS3}(c) and (f) illustrate the SOTI$^{\text{II}}$ phase, where the hinge modes now appear along the $x$ and $y$-directions, consistent with our earlier results with uniform coupling. These observations suggest that the stability of the HyOTP, SOTI$^{\rm (I)}$, and SOTI$^{\rm (II)}$ phases considering a realistic decaying profile of the altermagnetic exchange.
	
	Then, we explicitly investigate the current switching effect considering these exponentially decaying profile of the altermagnetic exchange couplings $J_{xy}$ and $J_{xz}$. We compute the differential conductance $\frac{dI^{\rm (H)}}{dV}(eV)$ mediated via hinge modes, employing Eq.\,(11) of the main text, 
	and attach leads at the top and bottom surfaces of TI (as shown in the inset of Fig.\,\ref{FigS3}(g)). We depict the zero bias conductance, $\frac{dI^{\rm (H)}}{dV}(eV=0)$ as a function of $J_{xz}^0$ in Fig.\,\ref{FigS3}(g) for various choices of $\zeta/a$ with $J_{xy}^0=3t$. We find that the conductance is not quantized throughout the SOTI$^{\rm (I)}$ phase, in sharp contrast to the uniform coupling case. For large localization length, $\zeta/a=8$, there is a tendancy towards the conductance quantization. However, importantly, the current-switching behaviour still persists as $\frac{dI^{\rm (H)}}{dV}(eV=0)\ne 0$ in the SOTI$^{\rm (I)}$ phase and $\frac{dI^{\rm (H)}}{dV}(eV=0)= 0$ in the SOTI$^{\rm (II)}$ phase, irrespective of the lifting of quantization.
	Hence, even without assuming a uniform altermagnetic strength across the bulk, a spatially varying (exponentially decaying) magnetic exchange profile can give rise to the fascinating current-switching phenomena mediated via hinge modes that we propose in the main text.

		\begin{center}
	\specialsection{\hskip 0.4cm Effect of disorder}\label{Sec:VII}
\end{center}

\begin{figure}[h]
	\centering
	\includegraphics[width=0.7\textwidth]{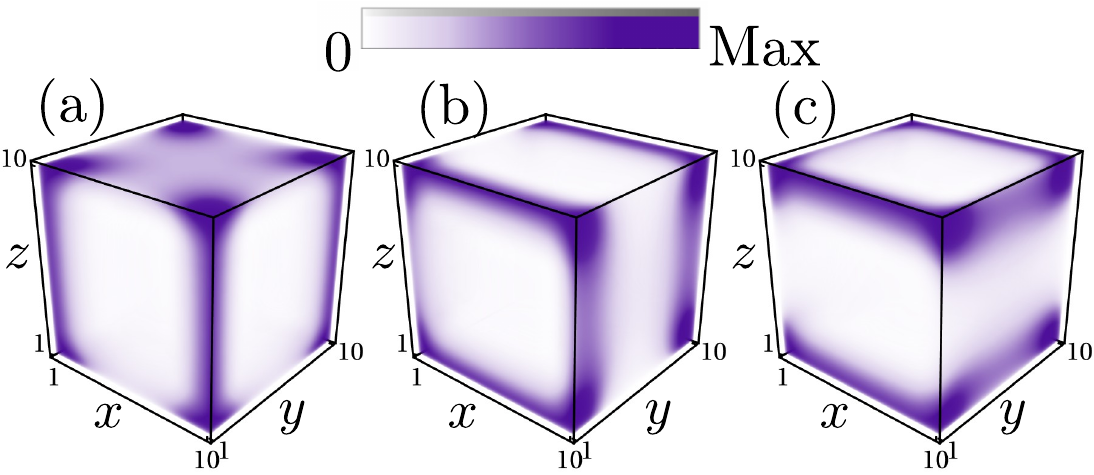}
	\caption{We present the disorder averaged LDOS at \(E = 0\) in the presence of random onsite disorder. In panel (a), the parameters are set to \(J_{xz} = 0\) and \(J_{xy} = 3t\). Panels (b) and (c) correspond to \(J_{xz} = 2.5t,\, J_{xy} = 3t\) and \(J_{xz} = 3t,\, J_{xy} = 2.5t\), respectively. The remaining model parameters are chosen as \(V_{\text{dis}} = 0.5t\), \(m_0 = 2t\), \(\lambda = t\), and \(t = 1\), on a \(10 \times 10 \times 10\) lattice. All results are averaged over 30 independent disorder configurations.
	}
	\label{FigS4}
\end{figure}

\begin{figure}[h]
	\centering
	\includegraphics[width=0.5\textwidth]{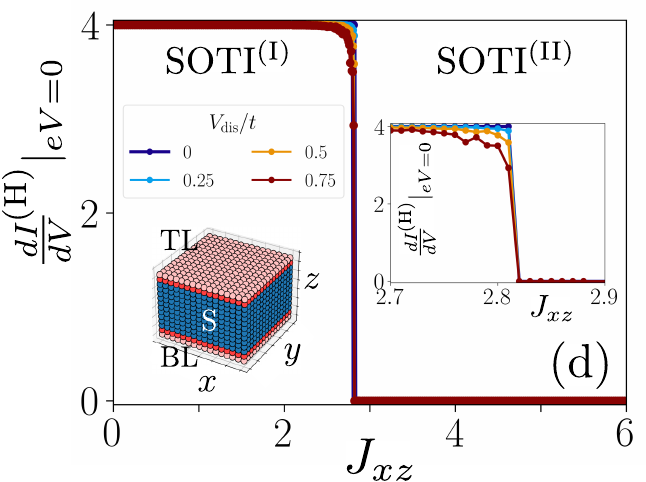}
	\caption{Variation of $\frac{dI^{\rm (H)}}{dV}|_{eV=0}$ is displayed as a function of $J_{xz}$ for various disorder strengths $V_{\rm dis}$ after averaging over 30 independent disorder configurations. In the inset, $\frac{dI^{\rm (H)}}{dV}|_{eV=0}$ close to the phase transition point is highlighted for better clarity. Other model parameters are chosen as: $(m_0,\lambda,J_{xy})=(t,t,3t)$ and $(L_x,L_y,L_z)=(20, 20,10)$.}
	\label{FigS4_Disorder}
\end{figure}

In the main text, we have discussed the stability of the differential conductance in presence of random onsite disorder. Here, we discuss the disorder effect in more detail. We add a disorder term to the main Hamiltonian, $\mc{H}(\mbf{k})$ (see Eq.\,(1) of the main text) given by, $\mc{H}_{\rm dis} = \sum_{\mbf{r}} V_{\mbf{r}} \,\mu_0\sigma_0s_0$, where $V_{\mbf{r}}$ is a random variable generated from a box distribution $[ -V_{\rm dis}, V_{\rm dis} ]$ and corresponds to the onsite potential at lattice site `$\mbf{r}=(x,y,z)$'. 

First,	we demonstrate that in all the three phases-namely the HyOTP, SOTI$^{\text{I}}$, and SOTI$^{\text{II}}$-both the hinge modes and surface states remain robust against weak disorder by investing the LDOS profiles. As shown in Fig.~\ref{FigS4}(a), the coexistence of surface and hinge modes persists even in the presence of onsite disorder. Furthermore, Figs.~\ref{FigS4}(b) and \ref{FigS4}(c), corresponding LDOS for the SOTI$^{\text{I}}$ and SOTI$^{\text{II}}$ phases respectively, exhibit that the characteristic feature of 1D hinge modes is preserved. Although the hinge modes become slightly more dispersive compared to the clean case, they continue to exist, indicating that weak to moderate disorder does not destroy the underlying topological features of the system.

Then, we discuss the stability of our transport results against disorder. Using the lead configuration shown in the inset of Fig.\,\ref{FigS4_Disorder}, we compute the zero-bias differential conductance mediated via the hinge modes, $\frac{dI^{\rm (H)}}{dV}|_{eV=0}$ (see Eq.\,(11) of the main text). The results are obtained after averaging over 30 independent disorder realizations. In Fig.\,\ref{FigS4_Disorder}, we show $\frac{dI^{\rm (H)}}{dV}|_{eV=0}$ 
as a function of $J_{xz}$ choosing different disorder strengths.

We find that close to the transition between the SOTI$^{\rm (I)}$ and SOTI$^{\rm (II)}$ phases, the conductance quantization is lifted in presence of moderate disorder strength (0.5t). In contrast, deep inside the SOTI$^{\rm (I)}$ phase, the quantization remains intact. This behavior can be understood as follows: near the transition point, the bulk gap becomes small, allowing disorder-induced scattering of hinge states into bulk states. However, deep inside the SOTI$^{\rm (I)}$ phase, the larger bulk gap enhances the protection of the hinge modes, thereby preserving conductance quantization. Importantly, irrespective of the quantization breakdown, current switching behavior-the central proposal of our work remains clearly visible even with larger disorder strength (0.75t). Therefore, our results establish that the proposed switching mechanism is robust against scalar random disorder and remains experimentally viable, rather than being a clean-limit phenomenon.




		\begin{center}
	\specialsection{\hskip 0.4cm Third-order TI: a possibility}\label{Sec:VIII}
\end{center}
In this section, we put forward a purely theoretical proposal for engineering third order TI (TOTI). 
If a unconventional magnet with $d$-wave magnetic order exists in the form of $d_{r^2-3z^2}$ and $r^2=x^2+y^2+z^2$, then the form of the exchange term can be 
given as $J_2(k_x^2+k_y^2-2k_z^2)$.

	\begin{center}
	\specialsubsection{\hskip 0.4cm Bulk Hamiltonian}
\end{center}

We first rewrite the tight-binding Hamiltonian (Eq.~(\ref{eq:Hamiltonian})) as a low energy continuum model by replacing $\sin k_i \rightarrow k_i$ and $\cos k_i \rightarrow (1- k_i^2/2)$ with $i=x,y,z$ as,
\begin{equation}
	\begin{aligned}
		H=2\lambda \sum_{j=1}^3 k_j \Gamma_j+\left(m_0-t\sum_{j=1}^3 k_j^2\right) \Gamma_4 -\frac{J_{xy}}{2} \left(k_x^2-k_y^2\right)\Gamma_5+\frac{J_{xz}}{2} \left(k_z^2-k_x^2\right)\Gamma_5
		+J_2(k_x^2+k_y^2-2k_z^2)\Gamma_6\ ,
	\end{aligned}
\end{equation}
where, $\Gamma_6=\mu_x\sigma_0s_x$.
Employing OBC in all three spatial directions, the hinge states become fully gapped (see Fig.~\ref{Fig5}(a-c)), giving rise to a TOTI phase. In this regime, the octupolar winding number is quantized at $O_{xyz}=1$ as shown in  Fig.~\ref{Fig5}(d). This signifies a single zero-energy mode bound to each corner (see Fig.~\ref{Fig5}(f)). Finite-size scaling of the topological gap, performed for lattice sizes from $8\times8\times8$ to $16\times16\times16$, exhibits the gap closing systematically with increasing system size [inset of Fig.~\ref{Fig5}(d)]. For the $16\times16\times16$ lattice, the eigenvalue spectrum (Fig.~\ref{Fig5}(e)) reveals eight zero-energy modes, consistent with the presence of one mode per corner. The corresponding zero-energy LDOS distribution confirms sharply localized corner states as depicted in Fig.~\ref{Fig5}(f).
\begin{figure}[h]
	\centering
	\includegraphics[width=0.8\textwidth]{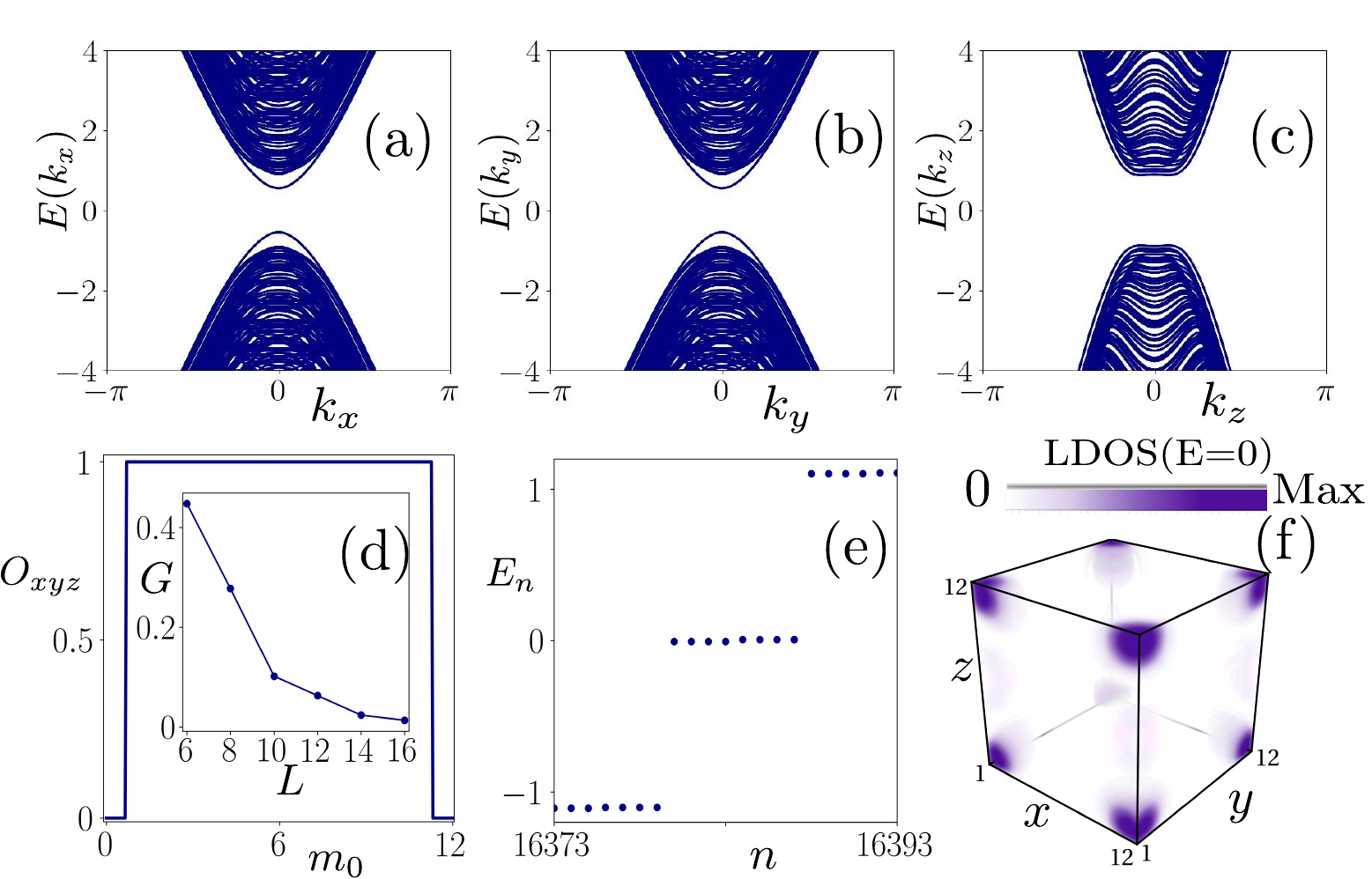}
	\caption{Panels (a-c) display the gapped hinge spectrum. The octupolar winding number $O_{xyz}$ is shown as a function of $m_{0}$ in panel (d). The inset shows the finite-size scaling of the energy gap at fixed $m_0 = 2t$. In panel (e), we depict the eigenvalue spectrum $E_{n}$ as a function of the state index $n$ considering $16\times 16 \times 16$ lattice size. Panel (f) illustrates the zero-energy LDOS, revealing corner localized modes in $12\times 12 \times 12$ system size. The model parameters are chosen as $J_{xy} = 2t$, $J_{xz}=0$ and $J_{2} = 4t$, while the remaining settings remain the same as mentioned in Fig.~2 of the main text.}
	\label{Fig5}
\end{figure}
	\begin{center}
	\specialsubsection{\hskip 0.4cm Topological Invariant}
\end{center}
To calculate the octapolar winding number ($O_{xyz}$), we instead use the chiral operator $U = \mu_y \sigma_0 s_0$. The corresponding octapole operator is defined as~ \cite{Benalcazar_2022} 
\begin{equation}
	O = \exp\left( -\frac{i 2\pi x y z}{L_x L_y L_z} \right)\ .
\end{equation}  

Projecting into the sublattice sectors yields
\begin{equation}
	\bar{O}_{A,B} = U_{A,B}^\dagger O U_{A,B}\ ,
\end{equation}  
and the octapolar winding number follows as  
\begin{equation}
	O_{xyz} = \frac{1}{2\pi i}\, \operatorname{Tr}\, \log \!\left( \bar{O}_{x,y,z}^A  \bar{O}_{x,y,z}^{B\dagger} \right)\ .
\end{equation} 

	\begin{center}
	\specialsubsection{\hskip 0.4cm Surface Hamiltonians}
\end{center}

To derive the surface Hamiltonian for the $xy$ surface, we consider OBC in the $z$ direction and PBC in $x$ and $y$ directions. Replacing $k_z \rightarrow -i\partial_z$ and keeping terms upto first order in $k_x$ and $k_y$, we divide the Hamiltonian into two parts:
\begin{equation}
	\begin{aligned}
		H_I&=(m_0+t\partial_z^2)\Gamma_4-2i\lambda\partial_z\Gamma_3\ ,\\
		H_{II}&=2\lambda k_x\Gamma_1+2\lambda k_y \Gamma_2 +\frac{J_{xz}}{2}\partial_z^2\Gamma_5+J_2\partial_z^2\Gamma_6\ .
	\end{aligned}
\end{equation}	

Solving the Hamiltonian as described before we obtain the surface Hamiltonian as
\begin{equation}
	H_{xy}^S=-2\lambda k_x\sigma_xs_y-2\lambda k_y\sigma_zs_0-\frac{J_2m_0}{t}\sigma_xs_x+\frac{J_{xz}m_0}{2t}\sigma_xs_z\ ,
\end{equation}

For the $yz$ surface, we impose OBC along $x$ and PBC along $y$ and $z$. The surface Hamiltonian becomes
\begin{equation}
	H_{yz}^S=-2\lambda k_y\sigma_zs_0+2\lambda k_z\sigma_xs_y +\frac{J_2m_0}{2t}\sigma_xs_x+\frac{(J_{xy}-J_{xz})m_0}{2t}\sigma_xs_z\ ,
\end{equation}

Similarly, for the $xz$ surface, we impose OBC along $y$, while $x$ and $z$ have PBC. 
The surface Hamiltonian can be written as
\begin{equation}
	H_{xz}^S=-2\lambda k_x\sigma_zs_0+2\lambda k_z\sigma_xs_0-\frac{J_2m_0}{2t}\sigma_ys_z+\frac{J_{xy}m_0}{2t}\sigma_ys_x\ .
\end{equation}

Using the surface Hamiltonians $H_{xy}^S, H_{yz}^S, H_{xz}^S$, we proceed to construct the Hinge theory. 

	\begin{center}
	\specialsubsection{\hskip 0.4cm Hinge Theory}
\end{center}

{\underline{\textbf{Hinge mode of $xy$ along $y$}}}:

To obtain the hinge modes, we start from the surface Hamiltonian. Considering OBC along $x$ and PBC along $y$, the Hamiltonian decomposes into:
\begin{equation}
	\begin{aligned}
		H_I&= 2i\lambda \partial_x  \sigma_xs_y +\frac{J_{xz}m_0}{2t}\sigma_xs_z\ ,\\
		H_{II}&= -2\lambda k_y  \sigma_zs_0 -\frac{J_{2}m_0}{t}\sigma_xs_x\ .
	\end{aligned}	
\end{equation}
We solve $H_I|\psi\rangle=0$ with boundary condition $|\psi\rangle \to 0$ as $x \to 0$. 
The solution is:
\begin{equation}
	|\psi\rangle \sim e^{-\zeta x+ik_yy} |\chi_{\alpha} \rangle\ ,
\end{equation}
with 
\[
\zeta=\left\{-\frac{J_{xz}m_0}{4\lambda t},-\frac{J_{xz}m_0}{4\lambda t},\frac{J_{xz}m_0}{4\lambda t},\frac{J_{xz}m_0}{4\lambda t}\right\},
\]
and
\begin{equation}
	\chi_{1}=
	\begin{pmatrix}
		0\\0\\1\\1
	\end{pmatrix}, \quad 
	\chi_{2}=
	\begin{pmatrix}
		1\\1\\0\\0
	\end{pmatrix}.
\end{equation}
The hinge Hamiltonian along $x$ is:
\begin{equation}
	H_{xy,y}^H=2\lambda k_y\sigma_z-\frac{J_2m_0}{t}\sigma_x\ .
\end{equation}

{\underline{\textbf{Hinge mode of $xy$ along $x$}}}:
\vspace{+0.2cm}

For hinge mode along $x$, we consider OBC along $y$ and PBC along $x$, the Hamiltonian decomposes into:
\begin{equation}
	\begin{aligned}
		H_I&= 2i\lambda \partial_y  \sigma_zs_0 +\frac{J_{xz}m_0}{2t}\sigma_xs_z\ ,\\
		H_{II}&= -2\lambda k_x  \sigma_xs_y -\frac{J_{2}m_0}{t}\sigma_xs_x\ .
	\end{aligned}	
\end{equation}

The zero-energy solution is of the form
\begin{equation}
	|\psi\rangle \sim e^{-\zeta y+ik_xx} |\chi_{\alpha}\rangle\ ,
\end{equation}
with the same $\zeta$ as mentioned before and
\begin{equation}
	\chi_{1}=
	\begin{pmatrix}
		0\\i\\0\\1
	\end{pmatrix}, \quad 
	\chi_{2}=
	\begin{pmatrix}
		-i\\0\\1\\0
	\end{pmatrix}\ .
\end{equation}

The hinge Hamiltonian becomes
\begin{equation}
	H_{xy,x}^H=-2\lambda k_x\sigma_x+\frac{J_2m_0}{t}\sigma_y\ .
\end{equation}

{\underline{\textbf{Hinge mode of $xz$ along $z$}}}:
\vspace{+0.2cm}

For the $xz$ surface and hinge modes along $z$, considering OBC along $x$ and PBC 
along $z$, we decompose:
\begin{equation}
	\begin{aligned}
		H_I&= 2i\lambda \partial_x  \sigma_zs_0 +\frac{J_{xy}m_0}{2t}\sigma_ys_x\ ,\\
		H_{II}&= 2\lambda k_z  \sigma_xs_0 -\frac{J_{2}m_0}{2t}\sigma_ys_z\ .
	\end{aligned}	
\end{equation}

The zero-energy solution is:
\begin{equation}
	|\psi\rangle \sim e^{-\zeta x+ik_zz} |\chi_{\alpha}\rangle\ ,
\end{equation}
with 
\[
\zeta=\left\{-\frac{J_{xy}m_0}{4\lambda t},-\frac{J_{xy}m_0}{4\lambda t},\frac{J_{xy}m_0}{4\lambda t},\frac{J_{xy}m_0}{4\lambda t}\right\},
\]

and
\begin{equation}
	\chi_{1}=
	\begin{pmatrix}
		-1\\0\\0\\1
	\end{pmatrix}, \quad 
	\chi_{2}=
	\begin{pmatrix}
		0\\-1\\1\\0
	\end{pmatrix}.
\end{equation}
The hinge Hamiltonian along $z$ can be written as
\begin{equation}
	H_{xz,z}^H=-2\lambda k_z\sigma_x+\frac{J_2m_0}{2t}\sigma_y\ .
\end{equation}

{\underline{\textbf{Hinge mode of $xz$ along $x$}}}:
\vspace{+0.2cm}

Employing OBC along $z$ and PBC along $x$, the zero-energy solution is:
\begin{equation}
	|\psi\rangle \sim e^{-\zeta z+ik_xx} |\chi_{\alpha}\rangle\ ,
\end{equation}
with
\begin{equation}
	\chi_{1}=
	\begin{pmatrix}
		i\\-i\\0\\0
	\end{pmatrix}, \quad 
	\chi_{2}=
	\begin{pmatrix}
		0\\0\\1\\1
	\end{pmatrix}.
\end{equation}
The hinge Hamiltonian is:
\begin{equation}
	H_{xz,x}^H=-2\lambda k_x\sigma_z+\frac{J_2m_0}{2t}\sigma_x\ .
\end{equation}

{\underline{\textbf{Hinge mode of $yz$ along $z$}}}:
\vspace{+0.2cm}

For the $yz$ surface, with OBC along $y$ and PBC along $z$, we write:
\begin{equation}
	\begin{aligned}
		H_I&= 2i\lambda \partial_z \sigma_zs_0 +\frac{J_{xy}m_0}{2t}\sigma_xs_z\ ,\\
		H_{II}&= 2\lambda k_z  \sigma_xs_y +\frac{J_{2}m_0}{2t}\sigma_xs_x-\frac{J_{xz}m_0}{2t}\sigma_xs_z\ .
	\end{aligned}	
\end{equation}
The solution becomes
\begin{equation}
	|\psi\rangle \sim e^{-\zeta y+ik_zz} |\chi_{\alpha}\rangle\ ,
\end{equation}
with 
\[
\zeta=\left\{-\frac{J_{xy}m_0}{4\lambda t},-\frac{J_{xy}m_0}{4\lambda t},\frac{J_{xy}m_0}{4\lambda t},\frac{J_{xy}m_0}{4\lambda t}\right\},
\] 
and
\begin{equation}
	\chi_{1}=
	\begin{pmatrix}
		0\\i\\0\\1
	\end{pmatrix}, \quad 
	\chi_{2}=
	\begin{pmatrix}
		-i\\0\\1\\0
	\end{pmatrix}\ .
\end{equation}
The hinge Hamiltonian can be written as
\begin{equation}
	H_{yz,z}^H=2\lambda k_z\sigma_x-\frac{J_2m_0}{2t}\sigma_y\ .
\end{equation}

{\underline{\textbf{Hinge mode of $yz$ along $y$}}}:
\vspace{+0.2cm}

In this case, imposing OBC along $z$ and PBC along $y$, the solution is:
\begin{equation}
	|\psi\rangle \sim e^{-\zeta z+ik_yy} |\chi_{\alpha}\rangle\ ,
\end{equation}
with
\begin{equation}
	\chi_{1}=
	\begin{pmatrix}
		0\\0\\-1\\1
	\end{pmatrix}, \quad 
	\chi_{2}=
	\begin{pmatrix}
		-1\\1\\0\\0
	\end{pmatrix}.
\end{equation}
The hinge Hamiltonian is:
\begin{equation}
	H_{yz,y}^H=2\lambda k_y\sigma_z-\frac{J_2m_0}{2t}\sigma_x\ .
\end{equation}

	\begin{center}
	\specialsubsection{\hskip 0.4cm Corner mode solution}
\end{center}
We now provide the solution for the corner modes located at $x=y=z=0$.  
Solving the hinge Hamiltonian with appropriate boundary conditions and matching at $x=y=z=0$, we obtain
\begin{equation}
	\begin{aligned}
		\Phi& \sim c_1^x\phi_1e^{\frac{-J_2m_0x}{2 \lambda t}}+c_2^x\phi_2e^{\frac{-J_2m_0x}{4\lambda t}} \quad : \text{along $x$},\\
		&\sim c_1^y\phi_2e^{\frac{-3 J_2m_0y}{4\lambda t}}\quad : \text{along $y$},\\
		&\sim c_1^z\phi_1e^{\frac{-J_2 m_0 z}{2 \lambda t}}\quad : \text{along $z$},
	\end{aligned}
\end{equation}
where the spinors are 
\[
\phi_1=\{1,0\}, \quad \phi_2=\{-i,1\}.
\]
Therefore, the corner modes decay along all the directions ($x$, $y$, $z$) with different localizations lengths depending on the parameters $t, \lambda, m_{0}$ and $J_{2}$. 

\end{onecolumngrid}	

\end{document}